\documentclass[11pt]{article}
\pdfoutput=1
\usepackage{jheppub}
\usepackage{graphicx}
\usepackage{amssymb}
\usepackage{amsmath,amssymb}
\usepackage{slashed}
\usepackage{soul}
\usepackage{hyperref}
\usepackage{caption}
\usepackage{xcolor}
\usepackage{dsfont}
\usepackage{verbatim}
\usepackage{subfig}
\usepackage{amsmath,amssymb,amscd,amsfonts,mathtools}
\usepackage{xifthen}
\usepackage{xcolor}
\definecolor{markgreen}{RGB}{230,243,230}
\definecolor{darkolivegreen}{rgb}{0.33, 0.42, 0.18}
\definecolor{darkpastelgreen}{rgb}{0.01, 0.75, 0.24}

\usepackage[toc,page]{appendix}
\usepackage{epsfig}
\usepackage{epstopdf}
\usepackage{latexsym}
\usepackage{graphicx}
\usepackage{booktabs}
\usepackage{bbm}
\usepackage{color}
\usepackage{physics}
\usepackage{tensor}
\usepackage{verbatim}
\usepackage{subfig}
\usepackage{tikz}
\usepackage{ifthen}
\usepackage{array}
\usetikzlibrary{matrix}
\usetikzlibrary{decorations.markings,calc,shapes,decorations.pathmorphing,patterns,decorations.pathreplacing,arrows.meta}
\usetikzlibrary{positioning}
\usepackage{hyperref}
\usetikzlibrary{patterns}
\usetikzlibrary{positioning}
\newdimen\mydim
\def\beq{\begin{equation}}
\def\eeq{\end{equation}}
\def\beqa{\begin{eqnarray}}
\def\eeqa{\end{eqnarray}}
\def\NN{{\cal {N}}}
\def\IS{{\bf {S}}}
\def\IZ{{\bf {Z}}}

\pdfoutput=1
\makeatletter
\def\@fpheader{\relax}
\newcommand*{\ov}[1]{%
  $\m@th\overline{\mbox{#1}}$%
}
\newcommand*{\ovA}[1]{%
  $\m@th\overline{\mbox{#1}\raisebox{3mm}{}}$%
}
\newcommand*{\ovB}[1]{%
$\m@th\overline{\mbox{#1\rule{0pt}{3mm}}}$%
}
\newcommand*{\ovC}[1]{%
  $\m@th\overline{\mbox{#1\strut}}$%
}
\newcommand*{\ovD}[1]{%
  $\m@th\overline{\mbox{#1\vphantom{\"A}}}$%
}
\newcommand*{\ovE}[1]{%
  $\m@th\overline{\raisebox{0pt}[1.2\height]{#1}}$%
}
\newcommand*{\ovF}[1]{%
  $\m@th\overline{\raisebox{0pt}[\dimexpr\height+1mm\relax]{#1}}$%
}
\newcommand*{\ovG}[1]{%
  $\m@th\overline{\raisebox{0pt}[\dimexpr\height+1mm\relax]{#1\vphantom{A}}}$%
}
\makeatother
\newboolean{showcomments}
\setboolean{showcomments}{false}
\newcommand\rem[1]{\ifthenelse{\boolean{showcomments}}{{#1}}{}}
\newcommand{\be}{\begin{equation}}
\newcommand{\ee}{\end{equation}}

\usepackage{xifthen}
\newcommand{\dalembert}[1][]{\ifthenelse{\isempty{#1}}{\Box}{#1\Box}}
\usetikzlibrary{decorations.pathmorphing}
\tikzset{snake it/.style={decorate, decoration=snake}}
\title{\Large Wet Hair: Global Symmetries in Entanglement Islands}
\author{Hao Geng$^{a}$, Jes\'{u}s Huertas$^{b}$, Andreas Karch$^{c}$, Lisa Randall$^{a}$ and Dawson Thomas$^{a}$}
\affiliation{$^{a}$
Gravity, Spacetime, and Particle Physics (GRASP) Initiative, Harvard University, 17 Oxford St., Cambridge, MA, 02138, USA.}
\affiliation{$^{b}$ Instituto Balseiro, Centro Atómico Bariloche, S.C. de Bariloche, 8400, Río Negro, Argentina.}
\affiliation{$^{c}$ Theory Group, Department of Physics, University of Texas, Austin, TX 78712, USA.}
\emailAdd{haogeng@fas.harvard.edu,  j.huertas@csic.es, karcha@utexas.edu, randall@g.harvard.edu,dthomas@g.harvard.edu}

\abstract{A central conjecture in quantum gravity is the non-existence of global symmetries. As a fully unitary theory, there is no information loss in  a UV complete quantum gravity theory. We see both these concepts reflected in the AdS/CFT correspondence, which tells us that dynamical processes in AdS are fully captured by a manifestly unitary CFT with no information loss. Furthermore, global symmetries of the CFT are dual to gauge symmetries in the AdS, which implies no global symmetry in the AdS. In this work, we provide concrete evidence for the connection between the non-existence of global symmetries and the absence of information loss in quantum gravity.  We study the \textit{island setups} in which a gravitational AdS is coupled with a nongravitational bath on its boundary. In such theories, the information in the AdS can be lost to the bath.  We provide concrete examples with global symmetries in the island setup, from both the bottom-up and the top-down perspectives. We argue that these global symmetries are consistent due to \textit{entanglement islands}, in which holography is realized in a novel fashion. The global symmetries we construct are all mixed with spontaneously broken gauge symmetries. We will show that this fact has two implications: \textbf{1)} The black hole hair is detectable in the bath (``wet hair"); \textbf{2)} a resolution of a puzzle proposed by Harlow and Shaghoulian.}

\begin{document}
\maketitle
\flushbottom
\pagebreak

\section{Introduction}\label{sec:intro}

It has long been conjectured that global symmetries cannot exist in a UV complete quantum gravity theory \cite{Banks:1988yz,Giddings:1988cx,Dine:1992vx,Kallosh:1995hi,ArkaniHamed:2006dz,Banks:2010zn} and therefore symmetries should be either gauged or broken. This conjecture about the absence of global symmetry applies for both continuous and discrete global symmetries. In this paper, we focus on continuous global symmetries. The basic argument for the absence of this type of global symmetries makes use of the Bekenstein-Hawking entropy formula for a black hole
\begin{equation}
    S_{BH}=\frac{A}{4G_{N}}\,,
\end{equation}
and the no-hair theorem of black holes. The Bekenstein-Hawking formula suggests that a black hole has a finite number of microstates upper bounded by $e^{S_{BH}}$ and the no-hair theorem states that black hole solutions are independent of the global symmetry charge inside the black hole. The latter suggests that coarse-grained properties of black holes are not affected by adding any amount of global symmetry charges, which however is in tension with the former as it suggests that there would exist an infinite number of black hole microstates distinguished by their global symmetry charges. Hence, global symmetries had better not exist in quantum gravity. We note that within AdS/CFT, a putative proof of the absence of global symmetries has been put forward in \cite{Harlow:2018jwu,Harlow:2018tng} (see also \cite{Chen:2020ojn,Hsin:2020mfa,Bah:2022uyz} for some other attempts from a rather different perspective).

The finite Hilbert space the logarithm of whose dimension scales linearly with the black hole horizon area is the early motivation for the holographic principle \cite{tHooft:1993dmi, Susskind:1994vu}. This principle is explicitly realized by the AdS/CFT correspondence \cite{Maldacena:1997re,Gubser:1998bc,Witten:1998qj}, which states that quantum gravitational dynamics in an asymptotically anti-de Sitter (AdS) spacetime is dual to a conformal field theory (CFT) living on its asymptotic boundary. In the context of a black hole in AdS space, the quantum mechanical system that describes its physics is the dual boundary CFT, which is manifestly unitary. Hence, the Bekenstein-Hawking entropy formula  suggests that the black hole dynamics, or the quantum gravitational theory that describes the black hole, is unitary. We also have seen from the previous paragraph that the Bekenstein-Hawking entropy formula suggests the absence of global symmetries in quantum gravity. Therefore, it is natural to conjecture that the absence of global symmetry is tied to the unitarity of quantum gravity. 

In this paper, we provide concrete evidence for the above conjecture from a novel perspective. We study a system in which a gravitational asymptotically AdS spacetime is coupled to a nongravitational bath on its asymptotic boundary, so that the gravitational theory is no longer unitary. \footnote{In principle, one can violate unitarity for a finite time, in which case our arguments still apply but the global symmetry can exist only increasingly deeper in the bulk.} Such setups, with a coupling to a bath, contain entanglement islands, which have been important to recent developments in quantum gravity \cite{Almheiri:2019psf,Penington:2019npb,Geng:2020qvw,Chen:2020uac,Chen:2020hmv}, 
so we call them \textit{island setups} for convenience.  Notice that our definition of island setups does not include all theories with islands, but only those in which unitarity is violated by the coupling of the gravitational system to a bath.  The so-called Page curve of the black hole radiation can be computed in this setup, which suggests that the dynamics of the union of the gravitational AdS and the bath is unitary. However, in this system the dynamics in the gravitational AdS alone is not unitary due to the bath coupling. Thus, if global symmetries  exist in such systems, it would demonstrate the connection between unitarity and the absence of global symmetries. More precisely, one should expect that particles charged under global symmetries can propagate into the gravitational AdS from the bath. Nevertheless, one might naively think that this expectation violates our former considerations of black holes in AdS. We will show in this paper that this is not the case and the reason is the existence of entanglement islands. We will show how entanglement islands allow for consistency of global symmetries with the above considerations of black holes. 

We will furthermore provide explicit examples of global symmetries in the island setup from both  bottom-up holographic constructions and  top-down string theory constructions. With the general argument that global symmetries can exist in the island setup and various concrete examples, we prove the early conjecture that global symmetries can exist in  quantum gravitational systems with disordering \cite{Benini:2022hzx}. Moreover, our work provides firm evidence for the conjecture that unitarity is essential to the absence of global symmetries in quantum gravity. To be clear we are considering only theories in which the full system  preserves unitarity, but the AdS system on its own does not.

We note that Ref. \cite{Harlow:2020bee} proposed that unitarity and the absence of global symmetries in gravitational theories can be connected. We attack this question with different considerations and conclusions in specific examples. We should also mention the possibility that this connection might be violated in some low dimensional theories \cite{Harlow:2020bee}. We do not address this question here but focus on the role of unitarity in quantum gravity theories which have propagating gravitational degrees of freedom.

This paper is organized as follows. In Sec.~\ref{sec:general} we review the physics of entanglement islands and provide a general argument of why global symmetry can exist in the island setup.  In Sec.~\ref{sec:GlobalSymIsland} we provide an explicit construction of a global symmetry in the island setup. In Sec.~\ref{sec:KR} we study the holographic dual of the island setup, i.e. the Karch-Randall braneworld \cite{Karch:2000ct,Karch:2000gx} and we construct global symmetries in the dual island setup using holography. In Sec.~\ref{sec:string} we provide constructions of global symmetries for the dual island setup in the string theory realization of the Karch-Randall braneworld. In Sec.~\ref{sec:resolvingHS} we resolve a puzzle proposed by Harlow and Shaghoulian in \cite{Harlow:2020bee} where they argued that global symmetries are in tension with entanglement islands. We conclude our paper with discussions in Sec.~\ref{sec:conclusion}.

\section{Entanglement Islands and Global Symmetries}\label{sec:general}
In this section, we review the physics of entanglement islands and then provide a general argument for why the existence of an entanglement island can address the potential inconsistency between global symmetries and the Bekenstein-Hawking entropy formula.

\subsection{A Lightening Review of Entanglement Islands}

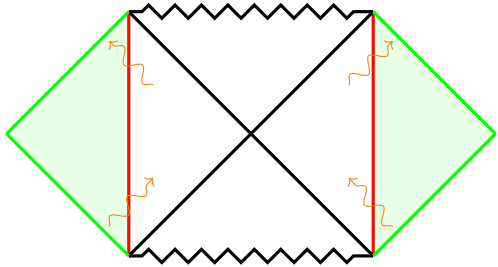
\begin{figure}[h]
    \centering
     \begin{tikzpicture}[scale=0.65,decoration=snake]
       \draw[-,very thick] 
       decorate[decoration={zigzag,pre=lineto,pre length=5pt,post=lineto,post length=5pt}] {(-2.5,0) to (2.5,0)};
       \draw[-,very thick,red] (-2.5,0) to (-2.5,-5);
       \draw[-,very thick,red] (2.5,0) to (2.5,-5);
         \draw[-,very thick] 
       decorate[decoration={zigzag,pre=lineto,pre length=5pt,post=lineto,post length=5pt}] {(-2.5,-5) to (2.5,-5)};
       \draw[-,very thick] (-2.5,0) to (2.5,-5);
       \draw[-,very thick] (2.5,0) to (-2.5,-5);
       \draw[-,very thick,green] (-2.5,0) to (-5,-2.5);
       \draw[-,very thick,green] (-5,-2.5) to (-2.5,-5);
        \draw[-,very thick,green] (2.5,0) to (5,-2.5);
       \draw[-,very thick,green] (5,-2.5) to (2.5,-5);
       \draw[fill=green, draw=none, fill opacity = 0.1] (-2.5,0) to (-5,-2.5) to (-2.5,-5) to (-2.5,0);
       \draw[fill=green, draw=none, fill opacity = 0.1] (2.5,0) to (5,-2.5) to (2.5,-5) to (2.5,0);
       \draw[->,decorate,orange] (-2.9,-4.4) to (-2,-3.4);
        \draw[->,decorate,orange] (-2,-1.5) to (-2.9,-0.6);
         \draw[->,decorate,orange] (2.9,-4.4) to (2,-3.4);
        \draw[->,decorate,orange] (2,-1.5) to (2.9,-0.6); 
    \end{tikzpicture}
    \caption{\small The Penrose diagram of an eternal black hole in AdS that is coupled with non-gravitational bath on its boundaries. The shaded green regions are the bath which consists of two half-Minkowski spaces. The asymptotic geometry of AdS allows a nice geometric picture of the coupling as gluing the bath to the black hole along their shared boundaries. The matter fields in the black hole spacetime obey transparent boundary conditions exchanging energy with the bath.}
    \label{pic:penroseisland}
\end{figure}

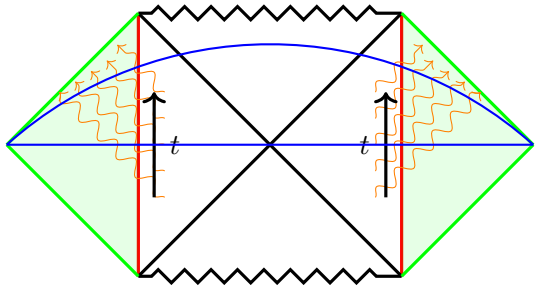
\begin{figure}[h]
    \centering
     \begin{tikzpicture}[scale=0.7,decoration=snake]
       \draw[-,very thick] 
       decorate[decoration={zigzag,pre=lineto,pre length=5pt,post=lineto,post length=5pt}] {(-2.5,0) to (2.5,0)};
       \draw[-,very thick,red] (-2.5,0) to (-2.5,-5);
       \draw[-,very thick,red] (2.5,0) to (2.5,-5);
         \draw[-,very thick] 
       decorate[decoration={zigzag,pre=lineto,pre length=5pt,post=lineto,post length=5pt}] {(-2.5,-5) to (2.5,-5)};
       \draw[-,very thick] (-2.5,0) to (2.5,-5);
       \draw[-,very thick] (2.5,0) to (-2.5,-5);
       \draw[-,very thick,green] (-2.5,0) to (-5,-2.5);
       \draw[-,very thick,green] (-5,-2.5) to (-2.5,-5);
        \draw[-,very thick,green] (2.5,0) to (5,-2.5);
       \draw[-,very thick,green] (5,-2.5) to (2.5,-5);
       \draw[fill=green, draw=none, fill opacity = 0.1] (-2.5,0) to (-5,-2.5) to (-2.5,-5) to (-2.5,0);
       \draw[fill=green, draw=none, fill opacity = 0.1] (2.5,0) to (5,-2.5) to (2.5,-5) to (2.5,0);
        \draw[->,decorate,orange] (-2,-1.5) to (-2.9,-0.6);
   \draw[->,decorate,orange] (-2,-2) to (-2.9-0.2,-1.1+0.2);
   \draw[->,decorate,orange] (-2,-2.5) to (-2.9-0.5,-1.6+0.5);
   \draw[->,decorate,orange] (-2,-3) to (-2.9-0.8,-2.1+0.8);
   \draw[->,decorate,orange] (-2,-3.5) to (-2.9-1.1,-2.6+1.1);
        \draw[->,decorate,orange] (2,-1.5) to (2.9,-0.6); 
           \draw[->,decorate,orange] (2,-2) to (2.9+0.2,-1.1+0.2);
   \draw[->,decorate,orange] (2,-2.5) to (2.9+0.5,-1.6+0.5);
   \draw[->,decorate,orange] (2,-3) to (2.9+0.8,-2.1+0.8);
   \draw[->,decorate,orange] (2,-3.5) to (2.9+1.1,-2.6+1.1);
         \draw[->,very thick,black] (-2.2,-3.5) to (-2.2,-1.5);
       \node at (-1.8,-2.5)
       {\textcolor{black}{$t$}};
        \draw[->,very thick,black] (2.2,-3.5) to (2.2,-1.5);
       \node at (1.8,-2.5)
       {\textcolor{black}{$t$}};
       \draw[-,thick, blue] (-5,-2.5) to (5,-2.5);
       \draw[-,thick,blue] (-5,-2.5) arc (90+41.81:90-41.81:7.5);
    \end{tikzpicture}
    \caption{\small One can choose a particular family of Cauchy slices (the blue slices) labeled by the time evolution in the diagram, such that more and more radiation from the black hole is captured by the Cauchy slice at later and later times.}
    \label{pic:penroseislandRad}
\end{figure}

Entanglement islands emerged from attempts to compute the Page curve of the radiation from a black hole. This was a hard question for both technical and conceptual reasons. At the conceptual level, an evaporating black hole's radiation and the black hole itself are living in the same spacetime, so it is unclear how to isolate the radiation from the black hole. At the technical level, it is unclear how the computation of fine-grained properties like the entanglement entropy can be done in a gravitational spacetime.  Thus, a natural setup that can circumvent the above difficulties is one with  a black hole spacetime coupled to  a non-gravitational bath by a transparent boundary condition for energy on its boundary. Black holes in AdS have a nice asymptotic structure  which allows the transparent boundary condition to be clearly modeled \cite{Porrati:2001gx,Porrati:2002dt,Porrati:2003sa,Duff:2004wh,Aharony:2006hz,Geng:2023ynk} (see Fig.\ref{pic:penroseisland}). In this setup, one can think of the non-gravitational bath as extracting the radiation from the black hole, so one can study the black hole radiation by studying this non-gravitational bath (see Fig.\ref{pic:penroseislandRad}).

\begin{figure}[h]
    \centering
\begin{tikzpicture}[scale=0.7]
    \draw[->,very thick,black] (-1,0) to (5,0);
    \draw[->,very thick,black] (0,-0.5) to (0,4);
    \node at (5.3,0) {\textcolor{black}{$t$}};
    \node at (0.3,4.2) {\textcolor{black}{$S(R)$}};
    \draw[-,very thick,red] (0,0.5) to (3,2.5);
    \draw[-,very thick,red] (3,2.5) to (5,2.5);
    \draw[-,dashed,very thick,black] (3,2.5) to (3,0);
    \node at (3,-0.3) {\textcolor{black}{$t_{p}$}};
\end{tikzpicture}  
\caption{\small The unitary Page curve. The initial growing is exponential and at the Page time $t_{p}$ the growing stops and $S(R)$ stays as a constant thereafter. }
    \label{pic:page}
\end{figure}
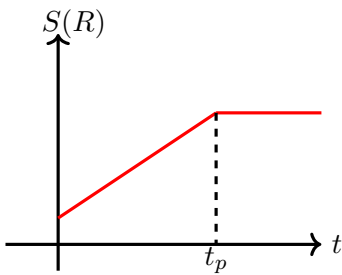

An interesting discovery from studying the above setup is the \textit{entanglement island}. It comes from the attempt to compute the entanglement entropy of the black hole radiation. The black hole radiation can be thought of as a subregion $R$ in the non-gravitational bath. The entanglement entropy of it can be computed using the \textit{quantum extremal surface} formula \cite{Engelhardt:2014gca}
\begin{equation}
    S(R)=\min_{ \mathcal{I}}\Big[S_{\text{QFT}}(R\cup\mathcal{I})+\frac{A(\partial\mathcal{I})}{4G_{N}}\Big]\,,\label{eq:islandformula}
\end{equation}
where $\mathcal{I}$ is a subregion in the gravitational AdS black hole spacetime, $S_{\text{QFT}}(R\cup \mathcal{I})$ is the entanglement entropy for the quantum fields in the subregion $R\cup \mathcal{I}$ as if it is in a fixed non-gravitational background, $A(\partial\mathcal{I})$ is the area of the boundary of the subregion $\mathcal{I}$ and $G_{N}$ is the AdS Newton's constant. In the above formula, the subregion $\mathcal{I}$ is called the entanglement island of $R$. This formula can be derived by using holography to translate the field theory replica path integral for $S(R)$ to a gravitational path integral in the full island setup and the \textit{replica wormhole} saddle in the resulting gravitational path integral proves Equ.~(\ref{eq:islandformula}) \cite{Almheiri:2019qdq,Penington:2019kki,Geng:2024xpj,Geng:2025efs}. The results of the formula Equ.~(\ref{eq:islandformula}) are the value of the entanglement entropy of the bath subregion $R$ and the location of its entanglement island $\mathcal{I}$ within the gravitational AdS black hole spacetime. The  dominance of the entanglement island  in the path integral  emerges at late time (see Fig.\ref{pic:penroseislandRadIsland}) and the resulting $S(R)$ obeys a unitary Page curve (see Fig.\ref{pic:page}).

Physically, one can think that the bath subregion $R$ is capturing more and more radiation from the black hole as time evolves (see Fig.\ref{pic:penroseislandRad}). Thus, its entanglement entropy is initially growing with time according to Hawking's calculation \cite{Hawking:1974sw} and this growth is saying that the entanglement between the black hole and the radiation it already emitted is becoming stronger. Nevertheless, a basic consistency from quantum mechanics is that the entanglement entropy between two quantum systems, i.e. the black hole and the radiation $R$, is upper bounded by the smallest Hilbert space dimension among these two quantum systems, i.e. 
\begin{equation}
    S(R)\leq \min\{\log d_{BH},\log d_{R}\}\,.
\end{equation}
In the above setup, $R$ is a half-infinite subregion in a nongravitational field theory system whose Hilbert space is infinitely large but the black hole Hilbert space is finite and given by the Bekenstein-Hawking entropy formula, i.e.
\begin{equation}
    \log d_{BH}= 2S_{BH}=\frac{A}{2G_{N}}\,,\label{eq:upperbound}
\end{equation}
where we have an additional factor of two because in the scenario depicted in Fig.\ref{pic:penroseisland} we are considering an eternal black hole which in fact contains two entangled black holes. As a result, the entanglement entropy $S(R)$ cannot grow forever and its growth has to be cut off at some instant by the value in Equ.~(\ref{eq:upperbound}). The emergence of entanglement island at late time in the application of the formula Equ.~(\ref{eq:islandformula}) justifies this expectation and results in a time-dependence of $S(R)$ that is consistent with quantum mechanics (see Fig.\ref{pic:page}). 

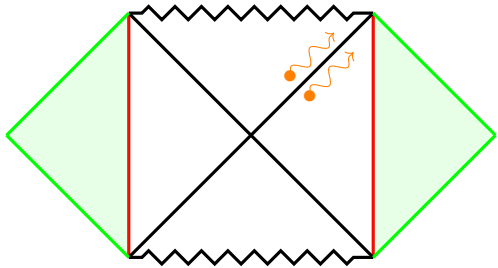
\begin{figure}[h]
    \centering
     \begin{tikzpicture}[scale=0.65,decoration=snake]
       \draw[-,very thick] 
       decorate[decoration={zigzag,pre=lineto,pre length=5pt,post=lineto,post length=5pt}] {(-2.5,0) to (2.5,0)};
       \draw[-,very thick,red] (-2.5,0) to (-2.5,-5);
       \draw[-,very thick,red] (2.5,0) to (2.5,-5);
         \draw[-,very thick] 
       decorate[decoration={zigzag,pre=lineto,pre length=5pt,post=lineto,post length=5pt}] {(-2.5,-5) to (2.5,-5)};
       \draw[-,very thick] (-2.5,0) to (2.5,-5);
       \draw[-,very thick] (2.5,0) to (-2.5,-5);
       \draw[-,very thick,green] (-2.5,0) to (-5,-2.5);
       \draw[-,very thick,green] (-5,-2.5) to (-2.5,-5);
        \draw[-,very thick,green] (2.5,0) to (5,-2.5);
       \draw[-,very thick,green] (5,-2.5) to (2.5,-5);
       \draw[fill=green, draw=none, fill opacity = 0.1] (-2.5,0) to (-5,-2.5) to (-2.5,-5) to (-2.5,0);
       \draw[fill=green, draw=none, fill opacity = 0.1] (2.5,0) to (5,-2.5) to (2.5,-5) to (2.5,0);
       \draw[->,decorate,orange] (1+0.2,1-0.2-2.5) to (1.9+0.2,1.9-0.2-2.5);
        \draw[->,decorate,orange] (1-0.2,1+0.2-2.5) to (1.9-0.2,1.9+0.2-2.5);
        \node at (1-0.2,1+0.2-2.5){\textcolor{orange}{$\bullet$}};
          \node at (1+0.2,1-0.2-2.5){\textcolor{orange}{$\bullet$}};
    \end{tikzpicture}
    \caption{\small An indication of the Hawking radiation. A virtual Hawking pair consists of two particles (the orange dots) with one inside the black hole horizon and one outside.}
    \label{pic:hawkingpair}
\end{figure}

However, a careful interpretation is required if one compares the above results with Hawking's calculation \cite{Hawking:1974sw}. Hawking's calculation can be interpreted heuristically as  virtual pairs of particles being  continuously produced near the black hole horizon, due to the lack of a global time translation invariance extending into that region, and each pair consists of two particles with one outside black hole horizon and one inside (see Fig.\ref{pic:hawkingpair}). The two particles in each pair are maximally entangled and the one that is outside the horizon is emitted as black hole radiation with the other one inside the horizon falling into the black hole. Thus, the entanglement between the emitted radiation and the black hole interior is growing (i.e. in Fig.\ref{pic:penroseislandRad} the region of the Cauchy slices behind the black hole horizon is growing exponentially with the time evolution). This monotonic growth superficially corresponds to the  size of the black hole interior which is expanding with time like an expanding universe.  This suggests that there is no violation of quantum mechanics at each instant. The trouble is caused by the Bekenstein-Hawking entropy formula which states that the system, i.e. the black hole, that is entangled with the emitted radiation
$R$ has a finite Hilbert space.

The holographic interpretation of an entanglement island is that the entanglement island $\mathcal{I}$ emerges at late time, which largely overlaps the black hole interior, and the physics in the entanglement island is fully encoded in the emitted radiation $R$. Thus, the quantum system that captures the dynamics of the black hole no longer captures all the physics in the black hole spacetime, including the interior of the black hole that overlaps with the island $\mathcal{I}$. This holographic interpretation can be phrased in terms of the entanglement wedge reconstruction in the AdS/CFT correspondence, which states that the entanglement wedge of the radiation region $R$ is $R\cup\mathcal{I}$, i.e. the physics in $R\cup\mathcal{I}$ in the island setup are fully captured by $R$ itself.\footnote{More precisely the physics in $R\cup\mathcal{I}$ that is fully captured by $R$ is the low energy physics in $R\cup I$ that doesn't significantly backreact on the geometry of the gravitational AdS black hole spacetime.} Similarly, the physics in the gravitational black hole spacetime that the quantum mechanical system for the black hole captures are only those in the complementary region of the island, i.e. $\bar{\mathcal{I}}$ which is the part of the blue intervals in Fig.\ref{pic:penroseislandRadIsland} inside the AdS black hole spacetime.  This significant change of the holographic picture circumvents the trouble we discussed at the end of the last paragraph.

Notice that entanglement islands universally exist in the island setup, even if there is no black hole in the gravitational AdS. This is because the island setups have dual field theory descriptions as a bath field theory with a boundary and in this description we can always consider a bipartition which splits this bath field theory system into a bigger subsystem and a smaller subsystem. We let the smaller subsystem contain the boundary. Thus, the entanglement entropy associated with this bipartition is upper bounded by the Hilbert space dimension of this smaller subsystem. To get this answer when we compute the entanglement entropy of the bigger subsystem in the island setup description, it is necessary that there is an entanglement island.

\begin{figure}[h]
    \centering
     \begin{tikzpicture}[scale=0.7,decoration=snake]
       \draw[-,very thick] 
       decorate[decoration={zigzag,pre=lineto,pre length=5pt,post=lineto,post length=5pt}] {(-2.5,0) to (2.5,0)};
       \draw[-,very thick,red] (-2.5,0) to (-2.5,-5);
       \draw[-,very thick,red] (2.5,0) to (2.5,-5);
         \draw[-,very thick] 
       decorate[decoration={zigzag,pre=lineto,pre length=5pt,post=lineto,post length=5pt}] {(-2.5,-5) to (2.5,-5)};
       \draw[-,very thick] (-2.5,0) to (2.5,-5);
       \draw[-,very thick] (2.5,0) to (-2.5,-5);
       \draw[-,very thick,green] (-2.5,0) to (-5,-2.5);
       \draw[-,very thick,green] (-5,-2.5) to (-2.5,-5);
        \draw[-,very thick,green] (2.5,0) to (5,-2.5);
       \draw[-,very thick,green] (5,-2.5) to (2.5,-5);
       \draw[fill=green, draw=none, fill opacity = 0.1] (-2.5,0) to (-5,-2.5) to (-2.5,-5) to (-2.5,0);
       \draw[fill=green, draw=none, fill opacity = 0.1] (2.5,0) to (5,-2.5) to (2.5,-5) to (2.5,0);
      \draw[->,decorate,orange] (-2,-1.5) to (-2.9,-0.6);
   \draw[->,decorate,orange] (-2,-2) to (-2.9-0.2,-1.1+0.2);
   \draw[->,decorate,orange] (-2,-2.5) to (-2.9-0.5,-1.6+0.5);
   \draw[->,decorate,orange] (-2,-3) to (-2.9-0.8,-2.1+0.8);
   \draw[->,decorate,orange] (-2,-3.5) to (-2.9-1.1,-2.6+1.1);
        \draw[->,decorate,orange] (2,-1.5) to (2.9,-0.6); 
           \draw[->,decorate,orange] (2,-2) to (2.9+0.2,-1.1+0.2);
   \draw[->,decorate,orange] (2,-2.5) to (2.9+0.5,-1.6+0.5);
   \draw[->,decorate,orange] (2,-3) to (2.9+0.8,-2.1+0.8);
   \draw[->,decorate,orange] (2,-3.5) to (2.9+1.1,-2.6+1.1);
         \draw[->,very thick,black] (-2.2,-3.5) to (-2.2,-1.5);
       \node at (-1.8,-2.5)
       {\textcolor{black}{$t$}};
        \draw[->,very thick,black] (2.2,-3.5) to (2.2,-1.5);
       \node at (1.8,-2.5)
       {\textcolor{black}{$t$}};
       \draw[-,thick,blue] (-5,-2.5) to (5,-2.5);
       \draw[-,thick,blue] (-5,-2.5) arc (90+41.81:90-41.81:7.5);
        \draw[-,very thick,red] (-5,-2.5) to (-3,-2.5);
       \draw[-,very thick,red] (5,-2.5) to (3,-2.5);
       \draw[-,very thick,red] (-5,-2.5) arc (90+41.81:115:7.5);
       \draw[-,very thick,red] (5,-2.5) arc (90-41.81:65:7.5);
 \node at (-3.8,-2.2)
       {\textcolor{red}{$R_{I}$}};
 \node at (3.8,-2.2)
       {\textcolor{red}{$R_{II}$}};
        \draw[-,very thick,purple] (0,-2.5+1.90983) arc (90:107:7.5);
          \draw[-,very thick,purple] (0,-2.5+1.90983) arc (90:73:7.5);
        \node at (0,-2.5+1.90983+0.3)
       {\textcolor{purple}{$\mathcal{I}$}};  
    \end{tikzpicture}
    \caption{\small One chooses a subregion $R=R_{I}\cup R_{II}$ (the red intervals) in the non-gravitational bath to model the emitted radiation from the black hole. This subregion is capturing more and more radiation from the black hole as time evolves. At late time entanglement island $\mathcal{I}$ (the purple interval) emerges.}
    \label{pic:penroseislandRadIsland}
\end{figure}
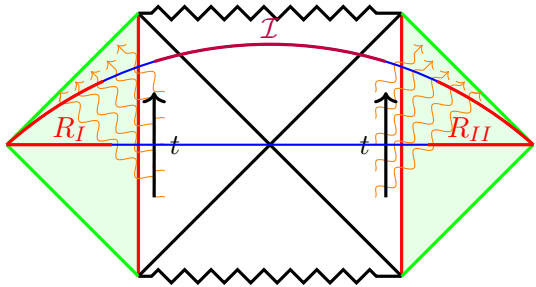

\subsection{The Consistency between Global Symmetries and Entanglement Islands}

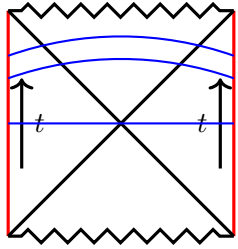
\begin{figure}[h]
    \centering
     \begin{tikzpicture}[scale=0.6,decoration=snake]
       \draw[-,very thick] 
       decorate[decoration={zigzag,pre=lineto,pre length=5pt,post=lineto,post length=5pt}] {(-2.5,0) to (2.5,0)};
       \draw[-,very thick,red] (-2.5,0) to (-2.5,-5);
       \draw[-,very thick,red] (2.5,0) to (2.5,-5);
         \draw[-,very thick] 
       decorate[decoration={zigzag,pre=lineto,pre length=5pt,post=lineto,post length=5pt}] {(-2.5,-5) to (2.5,-5)};
       \draw[-,very thick] (-2.5,0) to (2.5,-5);
       \draw[-,very thick] (2.5,0) to (-2.5,-5);
         \draw[->,very thick,black] (-2.2,-3.5) to (-2.2,-1.5);
       \node at (-1.8,-2.5)
       {\textcolor{black}{$t$}};
        \draw[->,very thick,black] (2.2,-3.5) to (2.2,-1.5);
       \node at (1.8,-2.5)
       {\textcolor{black}{$t$}};
       \draw[-,thick, blue] (-2.5,-2.5) to (2.5,-2.5);
       \draw[-,thick,blue] (-2.5,-2.5+1) arc (90+19.471:90-19.471:7.5);
          \draw[-,thick,blue] (-2.5,-2.5+1.5) arc (90+19.471:90-19.471:7.5);
    \end{tikzpicture}
    \caption{\small Cauchy slices (the blue slices) for black holes in standard gravity theories, i.e. without a bath. The size of the region behind the black hole horizon on the Cauchy slices is growing with time $t$ for an exponentially long period.}
    \label{pic:penroseCauchy}
\end{figure}

As we have discussed in Sec.~\ref{sec:intro}, the trouble with global symmetries in quantum gravity in the context of black holes is that the no-hair theorem suggests that black hole microstates can contain an arbitrary global symmetry charge. This is consistent with the fact that the black hole interior is very large as it is growing with time for an  exponentially long period (see the late time Cauchy slice in Fig.\ref{pic:penroseCauchy}) so it can support a large number of particles inside the black hole. Thus, one can start from a black hole microstate with low global symmetry charge, insert operators creating light charged particles outside the black hole and evolve the resulting state according to the time evolution as depicted in Fig.\ref{pic:penroseCauchy}. Then at late time the particles are swallowed by the the black hole resulting in higher-charged black hole microstates. These microstates are supported on late time Cauchy slices with particles charged under the global symmetry residing on the part of the Cauchy slices inside the black hole. 

This however contradicts the Bekenstein-Hawking entropy formula which suggests that the black hole is fully captured by a finite dimensional quantum system (i.e. the CFT) living in the asymptotic boundary of the AdS black hole spacetime. This consideration implies the absence of global symmetry in the standard gravitational theories. It is essential for the above argument that the dual quantum mechanical system captures all physics inside the black hole spacetime, which is guaranteed by the AdS/CFT correspondence.

\begin{figure}[h]
    \centering
     \begin{tikzpicture}[scale=0.6,decoration=snake]
       \draw[-,very thick,red] (-2.5,0) to (-2.5,-5);
       \draw[-,very thick,red] (2.5,0) to (2.5,-5);
       \draw[-,very thick,green] (-2.5,0) to (-5,-2.5);
       \draw[-,very thick,green] (-5,-2.5) to (-2.5,-5);
        \draw[-,very thick,green] (2.5,0) to (5,-2.5);
       \draw[-,very thick,green] (5,-2.5) to (2.5,-5);
       \draw[fill=green, draw=none, fill opacity = 0.1] (-2.5,0) to (-5,-2.5) to (-2.5,-5) to (-2.5,0);
       \draw[fill=green, draw=none, fill opacity = 0.1] (2.5,0) to (5,-2.5) to (2.5,-5) to (2.5,0);
        \draw[->,decorate,purple] (4.1,-2.5) to (3,-2);
    \end{tikzpicture}
    \caption{\small The dual field theory system to the setup in Fig.\ref{pic:penroseisland}. The dual contains two copies of the same field theory system living on a half-Minkowski space, i.e. with a boundary (the red vertical lines). The two field theories are in the thermal-field-double state. A particle-like excitation (the purple wave) can be created by a local operator insertion that propagates towards the boundary.}
    \label{pic:dualfield}
\end{figure}
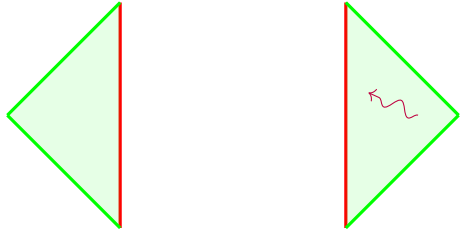

However, in the island setup the situation becomes quite different as the growing interior of the black hole is fully captured by a different quantum system, i.e. the bath subregion $R$, in addition to the original dual CFT that lives on the asymptotic boundary of the black hole spacetime. The bath subregion $R$ has an infinitely large Hilbert space so it is consistent for the interior of the black hole to contain a large number of light particles charged under a global symmetry with these particles supported on the late time Cauchy slice. Thus, we see that the existence of the entanglement island eliminates the tension between the existence of the global symmetry and the quantum mechanical description of the black hole. 

Furthermore, we now argue that in the island setup there can exist global symmetries or more precisely there can exist particles in the gravitational spacetime that are charged under global symmetries, that can be encoded and therefore detectable in the bath subregion $R$. We will provide explicit examples of global symmetries in the island setup of this type in the following sections.

\section{Global Symmetry in the island setup}\label{sec:GlobalSymIsland}
We now provide an explicit construction of a U(1) global symmetry in the island setup. This construction can be easily extended to non-Abelian continuous symmetries and potentially discrete symmetries. The basic input of this construction is that in the standard AdS/CFT correspondence, boundary CFT global symmetries are dual to bulk gauge symmetries with dynamical gauge fields. 

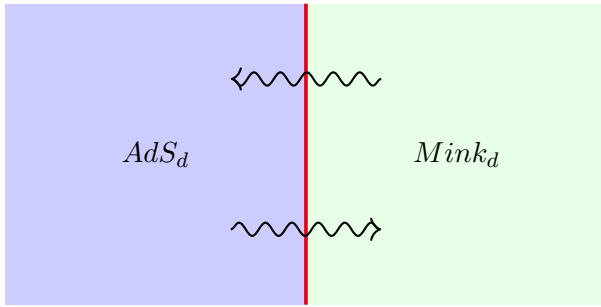
\begin{figure}[h] 
    \centering
    \begin{tikzpicture}
       \draw[-,very thick,red](0,-2) to (0,2);
       \draw[fill=green, draw=none, fill opacity = 0.1] (0,-2) to (4,-2) to (4,2) to (0,2);
           \draw[-,very thick,red](0,-2) to (0,2);
       \draw[fill=blue, draw=none, fill opacity = 0.2] (0,-2) to (-4,-2) to (-4,2) to (0,2);
       \node at (-2,0)
       {\textcolor{black}{$AdS_{d}$}};
        \node at (2,0)
       {\textcolor{black}{$Mink_{d}$}};
       \draw [-{Computer Modern Rightarrow[scale=1.25]},thick,decorate,decoration=snake] (-1,-1) -- (1,-1);
       \draw [-{Computer Modern Rightarrow[scale=1.25]},thick,decorate,decoration=snake] (1,1) -- (-1,1);
    \end{tikzpicture}
    \caption{We couple a gravitational AdS$_{d}$ universe (the blue shaded region) with a nongravitational bath (the green shaded region) by gluing them along the asymptotic boundary of the AdS$_{d}$ (the red vertical line). The nongravitational bath is modeled by another a half-Minkowski space which shares boundary with the AdS$_{d}$. We take the Poincar\'{e} coordinates in the AdS$_{d}$. In the dual field theory description, the coupling is achieved as described by Equ.~(\ref{eq:CFTdescription}).}
    \label{pic:islandmodel}
\end{figure}

We will consider a slightly simpler setup than the one in Sec.~\ref{sec:general}. We consider gravitational empty AdS coupled to a bath. Ref.\cite{Almheiri:2019psf,Penington:2019npb,Geng:2020qvw} showed that entanglement islands exist in such a setup. We consider an empty AdS$_{d}$ geometry, which has a dynamical U(1) gauge field $A_{\mu}(x,z)$ with a charged complex scalar field $\phi(x,z)$ having charge $q$ that leaks to a non-gravitational bath (see Fig.\ref{pic:islandmodel}). We take the non-gravitational bath to be a $d$-dimensional conformal field theory with an \textit{a priori} independent U(1) global symmetry that lives on  half-Minkowski space. We couple the AdS$_{d}$ and the bath such that the U(1) charge in the AdS can leak into the bath.\footnote{We emphasize that the AdS is gravitational but since we don't explicitly study gravitational dynamics in this paper besides taking into account of the existence of the island and using the AdS/CFT dictionary, we focus on the gauge field and charged matter sector.} The explicit descriptions of this setup are presented in the following subsections from both the dual field theory perspective and the point of view of the island setup itself.

\subsection{The Boundary Field Theory Analysis}\label{sec:CFTanalysis}

Here we consider the dual field theory description of our island setup, which is a CFT$_{d-1}$ at the boundary that is locally coupled to the CFT$_{d}$ of the bath on its boundary.  

This setup can be described by the following action 
\begin{equation}
    S_{\text{tot}}=S_{\text{CFT}_{d-1}}+S_{\text{CFT}_{d}}+g\int d^{d-1}x \Big(O_{1}(x)O_{2}(x)+O_{1}^{*}(x)O^{*}_{2}(x)\Big)\,.\label{eq:CFTdescription}
\end{equation}
In the above action we have two sectors that are coupled to each other. In the first sector, the CFT$_{d-1}$ is the holographic dual of the gravitational AdS$_{d}$, the U(1) gauge symmetry in the AdS$_{d}$ is dual to a U(1) global symmetry in the CFT$_{d-1}$ and $O_{1}(x)$ is the dual of the AdS charged scalar field $\phi_{1}(x,z)$ so it is a primary operator with charge $q$ under the U(1) global symmetry in CFT$_{d-1}$. In the second sector, we have the bath CFT$_{d}$ and $O_{2}(x)$ is the boundary extrapolation of a primary operator with charge $q'$ under the U(1) global symmetry of the CFT$_{d}$. The above two sectors are coupled to each other by the coupling term in Equ.~(\ref{eq:CFTdescription}) with the coupling strength $g$. This coupling allows the transfer of the U(1) charges between the two sectors and we will take it to be a marginal deformation. This simplifies our calculation as the CFT$_{d-1}$ is in the large $N$ limit so we can take the coupling to be exactly marginal to the leading order in perturbation theory.

We note that the coupling between the above two sectors explicitly breaks their respective U(1) global symmetries to a single preserved U(1) global symmetry. We will now argue that CFT$_{d-1}$ U(1) current, which is dual to the photon, is not conserved, and therefore it picks up an anomalous dimension. This will be important in the island setup as this anomalous dimension is  dual to the photon mass corresponding to the spontaneous breaking of the gauge symmetry.

Let's denote the current of the U(1) global symmetries in the CFT$_{d-1}$ and CFT$_{d}$ to be $J_{1}^{\mu}$ and $J_{2}^{a}$. Then the Noether theorem gives us
\begin{equation}
    \partial_{\mu}J_{1}^{\mu}(x)=-igq\Big(O_{1}(x)O_{2}(x)-O_{1}^{*}(x)O_{2}^{*}(x)\Big)\,,\label{eq:noether1}
\end{equation}
and
\begin{equation}
    \partial_{a} J_{2}^{a}(x,u)=-igq'\delta(u)\Big(O_{1}(x)O_{2}(x)-O_{1}^{*}(x)O_{2}^{*}(x)\Big)\,,\label{eq:noether2}
\end{equation}
where we use $u$ to denote the half-infinite coordinate in the bath due to the existence of the boundary at $u=0$. Thus, we have a conserved current
\begin{equation}
    J^{a}_{\text{cons}}(x,u)=q' \delta(u)J_{1}^{a}(x)-q J_{2}^{a}(x,u)\,,\label{eq:globalJ}
\end{equation}
where $J_{1}^{a}(x)=\delta^{a}_{\mu}J_{1}^{\mu}(x)$. This conserved current corresponds to an exact global U(1) symmetry which we will call U(1)$_{\text{global}}$ \cite{Karch:2023wui}. This global symmetry will persist in the dual island setup.

In the dual island setup, the gravitational AdS contains a gauge field $A_{a}(x,z)$ whose holographic dual is $J_{1}^{\mu}(x)$. From Equ.~(\ref{eq:noether1}) we can see that the bath coupling breaks this U(1) symmetry. Thus the current operator $J_{1}^{\mu}(x)$ is not protected from picking up an anomalous dimension. According to the AdS/CFT correspondence, this anomalous dimension will be dual to the mass of the gauge field $A_{a}(x,z)$. The leading order $g^2$ contribution to the anomalous dimension can be computed using the conformal algebra combined with the operator/state correspondence.\cite{Aharony:2006hz,Karch:2023wui}. The calculation goes as following.

We can use the state-operator correspondence in conformal field theory to construct a state
\begin{equation}
    \ket{\partial_{\mu}J_{1}^{\mu}}=P_{\mu}\ket{J^{\mu}_{1}}\,,
\end{equation}
which is created from the ground state by inserting the operator $\partial_{\mu}J_{1}^{\mu}$ in the infinite past. Here $P^{\mu}$ is the $(d-1)$-dimensional momentum operator. We can compute the inner product of this state with itself
\begin{equation}
    \bra{\partial_{\mu}J_{1}^{\mu}}\ket{\partial_{\nu}J_{1}^{\nu}}=\bra{J_{1}^{\mu}}K_{\mu}P_{\nu}\ket{J_{1}^{\nu}}\,,\label{eq:JJinner}
\end{equation}
where we used the fact that under the radial quantization the adjoint of the momentum operator $P^{\mu}$ is the special conformal transform $K^{\mu}$ \cite{Penedones:2016voo}. When the coupling is turned off the current $J_{1}^{\mu}(x)$ is a primary operator whose dual state $\ket{J_{1}^{\mu}}$ would thus be annihilated by the special conformal transform $K^{\mu}$. Hence, we can simplify Equ.~(\ref{eq:JJinner}) using the conformal symmetry algebra as
\begin{equation}
    \begin{split}
\bra{\partial_{\mu}J_{1}^{\mu}}\ket{\partial_{\nu}J_{1}^{\nu}}&=\bra{J_{1}^{\mu}}[K_{\mu},P_{\nu}]\ket{J_{1}^{\nu}}=2\bra{J_{1}^{\mu}}\eta_{\mu\nu}D-iM_{\mu\nu}\ket{J_{1}^{\nu}}\,,\\&=2C_{J_{1}}(d-1)\Big(\Delta_{J_1}-d+2\Big)\,,
    \end{split}
\end{equation}
where $D$ is the dilation operator, $\Delta_{J_1}$ is the conformal weight of the current operator $J_{1}^{\mu}(x)$, $C_{J_{1}}$ is a normalization coefficient, $M_{\mu\nu}$ is the generator of the Lorentz symmetry and we used
\begin{equation}
    iM_{\mu\nu}\ket{J_{1}^{\rho}}=-\eta_{\mu}^{\rho} \ket{J_{1\nu}}+\eta_{\nu}^{\rho}\ket{J_{1\mu}}\,,
\end{equation}
which follows from the fact that $J_{1}^{\nu}$ transforms as a vector under the Lorentz symmetry. Using Equ.~(\ref{eq:noether1}), we have 
\begin{equation}
\begin{split}
    2 C_{J_{1}}(d-1)(\Delta_{J_{1}}-d+2)=2g^{2}q^{2}\bra{O_{1}O_{2}}\ket{O_{1}O_{2}}\,.
    \end{split}
\end{equation}
Since we are doing perturbation theory only to leading order in the coupling\footnote{Note that the CFT can however be a strongly interacting system. We can nonetheless evaluate the two-point function because it is constrained by the conformal symmetry.} we have
\begin{equation}
\bra{O_{1}O_{2}}\ket{O_{1}O_{2}}=\bra{O_{1}}\ket{O_{1}}\bra{O_{2}}\ket{O_{2}}=C_{1}C_{2}\,,\label{eq:C1C2}
\end{equation}
where $C_{i}$'s are the two normalization constants with $i=1,2$. As a result, we have
\begin{equation}
    C_{J_{1}}(d-1)(\Delta_{J_{1}}-d+2)=g^{2}q^{2}C_{1}C_{2}\,.
\end{equation}
The normalization constants $C_{J_{1}}$, $C_{1}$ and $C_{2}$ are given by the two-point functions when the coupling is turned off
\begin{equation}
    \langle J_{1}^{\mu}(x)J_{1}^{\nu}(0)\rangle=C_{J_{1}}\frac{\eta^{\mu\nu}-\frac{x^{\mu}x^{\nu}}{ x^{2}}}{x^{2\Delta_{J_{1}}}}\,,\quad\langle O_{i}(x)O^{*}_{i}(0)\rangle=C_{i}\frac{1}{x^{2\Delta_{i}}}\,,
\end{equation}
where $\Delta_{i}$ is the conformal weight of the operator $O_{i}(x)$. Using the AdS/CFT correspondence, we expect the dual gauge boson mass in the gravitational AdS to be
\begin{equation}
    m^{2}_{\gamma}=(\Delta_{J_{1}}-1)(\Delta_{J_{1}}-d+2)=g^{2}q^{2}\frac{d-3}{d-1}\frac{C_{1}C_{2}}{C_{J_{1}}}+\mathcal{O}(g^{4})\,,\label{eq:photonmass}
\end{equation}
where we used $\Delta_{J_{1}}-1=d-3+\mathcal{O}(g^2)$ and the unit of $m_{\gamma}$ is set by $1/l_{AdS}$ if one restores the $l_{AdS}$ dependence. Note that the results so far, including the mass in Equ.~(\ref{eq:photonmass}), are independent of $q'$. We will see the importance of a nonzero $q'$ in the following subsection that it is nonetheless essential to how the particle inside the island can be charged under the global symmetry with the charge detectable in the bath.

\subsection{Spontaneous Symmetry Breaking in the Island Setup}
In the above subsection, we performed a dual boundary conformal field theory analysis of the island setup. In this subsection we study the gravitational island setup directly.  The field theory analysis and the AdS/CFT correspondence suggest that the gauge symmetry in the gravitational AdS part of the island setup is spontaneously broken and is mixed with an exact global U(1) symmetry that is supported on the union of the AdS and the bath. This global symmetry U(1) symmetry would be generated by Equ.~(\ref{eq:globalJ}). Thus, if true, a  global symmetry can exist in the island setup.

We show that the U(1) gauge symmetry in the AdS is spontaneously broken and the gauge boson becomes massive with a mass consistent with Equ.~(\ref{eq:photonmass}). We then show that the U(1) gauge symmetry is mixed with the exact global symmetry. Using these results, we demonstrate that the exact global U(1) symmetry charge of operators in the island can be detected in the bath. This is important in that it shows consistency with the holographic interpretation of the island to demonstrate how information in the island is  encoded in the bath.

\subsubsection{The Gauge Symmetry is Spontaneously Broken}

Before including the bath, let's first consider the relevant action in the AdS part of the island setup is 
\begin{equation}
\begin{split}
    S_{\text{\tiny AdS}}=-\int d^{d-1}xdz\sqrt{-g}\Big[g^{ab}D_{a}\phi_{1}(x,z)D^{*}_{b}\phi_{1}^{*}(x,z)&+m^{2}\phi_{1}(x,z)\phi_{1}^{*}(x,z)\Big]-\\&\frac{1}{4g_{e}^{2}}\int d^{d-1}x dz\sqrt{-g} F_{ab}F^{ab}\,,\label{eq:actionAdS}
    \end{split}
\end{equation}
where we have the covariant derivative $D_{a}=\partial_{a}-iqA_{a}(x,z)$ and $g_{e}^{2}$ is the coupling strength of the gauge field. We will use the Poincar\'{e} patch
\begin{equation}
    ds^{2}=g_{ab}dx^{a}dx^{b}=\frac{dz^{2}+\eta_{\mu\nu}dx^{\mu}dx^{\nu}}{z^{2}}\,.
\end{equation}
The action Equ.~(\ref{eq:actionAdS}) is invariant under the following gauge transform
\begin{equation}
    \phi_{1}(x,z)\rightarrow e^{iq\alpha(x,z)}\phi_{1}(x,z)\,,\quad A_{a}(x,z)\rightarrow A_{a}(x,z)+\partial_{a}\alpha(x,z)\,.
\end{equation}
The complex scalar field $\phi_{1}(x,z)$ obeys the transparent boundary condition near the asymptotic boundary $z\rightarrow0$, so the charge it carries leaks to the bath according to Equ.~(\ref{eq:CFTdescription}). The coupling terms in Equ.~(\ref{eq:CFTdescription}) can be thought of as double-trace deformations of the standard AdS/CFT setup. According to \cite{Witten:2001ua,Geng:2023ynk}, we have the near boundary expansion of the bulk scalar field $\phi_{1}(x,z)$
\begin{equation}
    \phi_{1}(x,z)= z^{\Delta} O_{1}(x) (1+\mathcal{O}(z^{2})) + z^{d-1-\Delta} \frac{g}{2\Delta-d+1}O_{2}^{*}(x)(1+\mathcal{O}(z^{2}))\,,\quad\text{as } z\rightarrow0\,,\label{eq:bctrans}
\end{equation}
where $\Delta=\frac{d-1}{2}+\sqrt{\frac{(d-1)^{2}}{4}+m^{2}}$ is the conformal weight of the dual operator $O_{1}(x)$. From the above, one can extract $\langle O_{1}(x)O_{1}^{*}(x)\rangle$ and $\langle O_{2}(x)O_{2}^{*}(x)\rangle$, i.e. $C_{1}$ and $C_{2}$ in Equ.~(\ref{eq:C1C2}), from the bulk Greens function of the canonically normalized field $\phi_{1}(x,z)$. The bulk Greens function is given by
\begin{equation}
    \langle T \phi_{1}(x_{1},z_{1})\phi^{*}_{1}(x_{2},z_{2})\rangle=\frac{1}{(2\Delta_{+}-d+1)^{2}+g^{2}}\Big[(2\Delta_{+}-d+1)^{2}G^{+}(X_{1},X_{2})+g^{2}G^{-}(X_{1},X_{2})\Big]\,,\label{eq:11}
\end{equation}
where we denote $X_{i}=(z_{i},x_{i})$ and we have the standard bulk AdS$_{d}$ Greens functions
\begin{equation}
    G^{\pm}(X_{1},X_{2})=\frac{\Gamma[\Delta_{\pm}]}{2\pi^{\frac{d-1}{2}}\Gamma[\Delta_{\pm}-\frac{d-3}{2}]}\frac{_{2}F_{1}[\frac{\Delta_{\pm}}{2},\frac{\Delta_{\pm}+1}{2},\Delta_{\pm}-\frac{d-3}{2},\frac{1}{Z^{2}}]}{(2Z)^{\Delta_{\pm}}}\,,\label{eq:G}
\end{equation}
with the invariant distance 
\begin{equation}
Z=\frac{z_{1}^{2}+z_{2}^{2}+\eta_{\mu\nu}(x_{1}-x_{2})^{\mu}(x_{1}-x_{2})^{\nu}}{2z_{1}z_{2}}\,, \label{eq:Z11}   
\end{equation} 
$\Delta_{+}=\Delta$ and $\Delta_{-}=d-1-\Delta$. Thus, we have
\begin{equation}
    C_{1}=\frac{(2\Delta-d+1)^2}{(2\Delta-d+1)^{2}+g^{2}}\frac{\Gamma[\Delta]}{2\pi^{\frac{d-1}{2}}\Gamma[\Delta-\frac{d-3}{2}]}\,,\quad C_{2}=\frac{(2\Delta-d+1)^2}{(2\Delta-d+1)^{2}+g^{2}}\frac{\Gamma[\Delta_{-}]}{2\pi^{\frac{d-1}{2}}\Gamma[\Delta_{-}-\frac{d-3}{2}]}\,,\label{eq:Cholographic}
\end{equation}
for which we used
\begin{equation}
    \lim_{Z\rightarrow\infty}\frac{_{2}F_{1}[\frac{\Delta_{\pm}}{2},\frac{\Delta_{\pm}+1}{2},\Delta_{\pm}-\frac{d-3}{2},\frac{1}{Z^{2}}]}{(2Z)^{\Delta_{\pm}}}=\frac{1}{(2Z)^{\Delta_{\pm}}}\,.
\end{equation}
Now we are ready to show that the U(1) gauge symmetry is spontaneously broken and the gauge boson $A_{\mu}(x,z)$ is massive. For this purpose, we first notice that, as opposed to the case of reflective boundary condition, the action Equ.~(\ref{eq:actionAdS}) is not invariant if one transforms the background gauge field
\begin{equation}
    A_{a}(x,z)\rightarrow A_{a}(x,z)+\partial_{a}\alpha(x,z)\,.\label{eq:gaugeA}
\end{equation}
This is due to the fact that under this transform we have
\begin{equation}
    \delta S=iq\int d^{d-1}x dz\sqrt{-g} \partial_{a}\alpha(x,z) J^{a}(x,z)+\mathcal{O}(\alpha^{2})\,,
\end{equation}
where we will ignore the $\mathcal{O}(\alpha^{2})$ terms  and we have the scalar current
\begin{equation}
    J^{a}(x,z)=g^{ab}\phi_{1}(x,z)D_{b}^{*}\phi_{1}^{*}(x,z)-g^{ab}D_{b}\phi_{1}(x,z) \phi_{1}^{*}(x,z)\,,
\end{equation}
which obeys the conservation law
\begin{equation}
    \partial_{a}J^{a}(x,z)=0\,,
\end{equation}
if one uses the equation of motion of $\phi_{1}(x,z)$. Thus, we have
\begin{equation}
    \delta S=iq \int d^{d-1}x \frac{1}{\epsilon^{d}}\alpha(x,\epsilon) J^{z}(x,\epsilon)\,,\label{eq:deltaS}
\end{equation}
where we regularized the asymptotic boundary of the AdS as $z=\epsilon$. The bulk gauge transform $\alpha(x,z)$ consists of large and small gauge transforms
\begin{equation}
\alpha(x,z)\sim \alpha_{\text{large}}(x) (1+\mathcal{O}(z^{2}))+\alpha_{\text{small}}(x) z^{d-1}(1+\mathcal{O}(z^{2}))\,,\quad\text{as }z\rightarrow0\,.    
\end{equation}
Thus, we can see from Equ.~(\ref{eq:deltaS}) that the large gauge transforms are broken due to the non-vanishing current flux according to the transparent boundary condition Equ.~(\ref{eq:bctrans}) but the small gauge transforms are preserved. This is consistent with the field theory expectation as bulk large gauge transforms (with a constant transformation parameter) are exactly the boundary global symmetry transform and the CFT$_{d-1}$ U(1) global symmetry generated by $J_{1}^{\mu}$ is broken due to the bath coupling. Moreover, from the AdS bulk point of view this observation supports the expectation that the U(1) gauge symmetry is spontaneously broken. This is because in the Higgs mechanism the small gauge symmetry is never broken and the broken symmetry is always the large gauge transform. In fact, the small gauge transforms are never broken as long as the current is locally conserved, though they can be nonlinearly realized in the spontaneously broken phase.

To show that the U(1) gauge symmetry is spontaneously broken and compute the gauge boson mass, let's restore the full U(1) gauge invariance by introducing a compensating field $\theta(x,z)$  \cite{PRESKILL1991323} which transforms nonlinearly under the gauge transform as
\begin{equation}
\theta(x,z)\rightarrow\theta(x,z)+q\alpha(x,z)\,.\label{eq:gaugetheta}
\end{equation}
 We will show that $\theta$ is a dynamical field  in the low energy theory in which the leaky scalar field has been integrated out. For this purpose, we we consider the manifestly gauge invariant AdS path integral
\begin{equation}
    Z_{\text{full}}=\int D[A]D[\phi_{1}]D[\theta]e^{iS_{\text{AdS}}[A,\phi_{1}]-i\int d^{d-1}xdz\sqrt{-g}\partial_{a}\theta(x,z)J^{a}(x,z)}\,,
\end{equation}
which is invariant under the U(1) gauge transform Equ.~(\ref{eq:gaugeA}) together with Equ.~(\ref{eq:gaugetheta}). We note that, for the case of the reflective boundary condition, the above operation is trivial, as the added $\theta(x,z)$ term in the action can be proved to be zero. In the transparent case, we can think of the action in Equ.~(\ref{eq:actionAdS}) as a gauge fixed version of the above full action with the gauge choice $\theta(x,z)=0$. We are interested in the low-energy effective action of the gauge field $A_{a}(x,z)$ and the compensating field $\theta(x,z)$ after integrating out the transparent charged matter field $\phi_{1}(x,z)$. This effective action is invariant under the gauge transform Equ.~(\ref{eq:gaugeA}) together with Equ.~(\ref{eq:gaugetheta}) so it can only be a functional of the invariant combination
\begin{equation}
    A_{a}(x,z)-\frac{1}{q}\partial_{a}\theta(x,z)\,.\label{eq:invariantcomb}
\end{equation}
We are interested in the leading quadratic terms. This can be computed by setting the gauge field $A_{a}(x,z)$ to zero and restoring it at the end using the invariant combination Equ.~(\ref{eq:invariantcomb}). Thus, we have
\begin{equation}
    S_{\text{eff},2}[\theta,0]=i\frac{1}{2}\int d^{d-1}xdz d^{d-1}x'dz' \sqrt{-g}\sqrt{-g'}\partial_{a}\theta(x,z)\langle j^{a}(x,z)j^{b}(x',z')\rangle\partial'_{b}\theta(x',z')\,,\label{eq:Seff}
\end{equation}
where we have
\begin{equation}
    j^{a}(x,z)=g^{ab}\phi_{1}(x,z)\partial_{b}\phi_{1}^{*}(x,z)-g^{ab}\partial_{b}\phi_{1}(x,z) \phi_{1}^{*}(x,z)\,.
\end{equation}
Since we are interested in the low-energy effective action, we  take the large distance limit for the two-point function $\langle j^{a}(x,z)j^{b}(x',z')\rangle$. This two-point function can be calculated using Wick-contraction
\begin{equation}
    \langle j^{a}(x,z)j^{b}(x',z')\rangle=2g^{ac}g'^{bd}\Big[\partial_{c}G(X,X')\partial'_{d}G(X,X')-G(X,X')\partial_{c}\partial'_{d}G(X,X')\Big]\,,\label{eq:jj}
\end{equation}
where $G(X_{1},X_{2})$ denotes the Greens function Equ.~(\ref{eq:11}). This Greens function is a single variable function of the invariant distance $Z$ in Equ.~(\ref{eq:Z11}) so we can denote it as $G(Z)$. Thus, we can simplify Equ.~(\ref{eq:jj}) as
\begin{equation}
    \langle j^{a}(x,z)j^{b}(x',z')\rangle=2g^{ac}g'^{bd}\Big[\Big(G'(Z)G'(Z)-G(Z)G''(Z)\Big)\partial_{c}Z\partial'_{d}Z-G(Z)G'(Z)\partial_{c}\partial'_{d}Z\Big]\,.\label{eq:jj2}
\end{equation}
To proceed, let us introduce the invariant tensors in AdS
\begin{equation}
    n_{a}=\frac{\partial_{a}Z}{\sqrt{Z^2-1}}\,,\quad n'_{b}=\frac{\partial_{b}'Z}{\sqrt{Z^2-1}}\,,\quad g_{ab'}=-\partial_{a}\partial_{b}'Z+\frac{\partial_{a}Z\partial_{b}'Z}{Z+1}\,,
\end{equation}
which obey $g_{ab'}n^{b'}=-n_{a}$ \cite{Allen:1985wd}. Then, the current-current two-point function Equ.~(\ref{eq:jj2}) can be decomposed as
\begin{equation}
     \langle j^{a}(x,z)j^{b}(x',z')\rangle=2g^{ac}g'^{bd}\Big[(Z^2-1)\Big(G'^2-GG''-\frac{GG'}{Z+1}\Big)n_{c}n'_{d}+G(Z)G'(Z)g_{cd'}\Big]\,.\label{eq:jj3}
\end{equation}
whose large distance limit, i.e. $Z\rightarrow\infty$, is given as
\begin{equation}
\begin{split}
     \langle j^{a}(x,z)j^{b}(x',z')\rangle=&
     2g^{ac}g'^{bd}\frac{4^{-1-\Delta_{-}}\pi^{1-d}\Delta_{-}\Gamma[\Delta_{-}]^{2}}{\Gamma[\Delta_{m}-\frac{d-3}{2}]Z^{2\Delta_{m}+1}}B^2(-g_{cd'}+n_{c}n'_{d})\\&-2g^{ac}g'^{bd}n_{c}n'_{d}\frac{(\Delta_{+}-\Delta_{-})^{2}\pi^{1-d}\Gamma[\Delta_{+}]\Gamma[\Delta_{-}]}{2^{d+2}\Gamma[\Delta_{+}-\frac{d-3}{2}]\Gamma[\Delta_{-}-\frac{d-3}{2}]Z^{d-1}}AB\,,\label{eq:jjlarged}
     \end{split}
\end{equation}
where $A=\frac{(2\Delta_{+}-d+1)^2}{(2\Delta_{+}-d+1)^2+g^2}$ and $B=\frac{g^2}{(2\Delta_{+}-d+1)^2+g^2}$. To the leading nontrivial order in the coupling $g$, the first row on the right-hand-side of Equ.~(\ref{eq:jjlarged}) can be ignored. Thus, we have the large distance and small coupling limit
\begin{equation}
\begin{split}
     \langle j^{a}(x,z)j^{b}(x',z')\rangle=-2g^{ac}g'^{bd}n_{c}n'_{d}\frac{(\Delta_{+}-\Delta_{-})^{2}\pi^{1-d}\Gamma[\Delta_{+}]\Gamma[\Delta_{-}]}{2^{d+2}\Gamma[\Delta_{+}-\frac{d-3}{2}]\Gamma[\Delta_{-}-\frac{d-3}{2}]Z^{d}}AB\,.\label{eq:jjlargedsmallg}
     \end{split}
\end{equation}
The Greens function of a massless scalar field with reflective boundary condition is given by
\begin{equation}
    G_{\text{massless}}(Z)=\frac{\Gamma[d-1]}{2\pi^{\frac{d-1}{2}}\Gamma[d-1-\frac{d-3}{2}]}\frac{_{2}F_{1}[\frac{d-1}{2},\frac{d}{2},d-1-\frac{d-3}{2},\frac{1}{Z^{2}}]}{(2Z)^{d-1}}\,,
\end{equation}
whose large distance limit is
\begin{equation}
     G_{\text{massless}}(Z)=\frac{1}{Z^{d-1}}\frac{2^{-d}\pi^{\frac{1-d}{2}}\Gamma[d-1]}{\Gamma[d-1-\frac{d-3}{2}]}\,,\quad\text{as }Z\rightarrow\infty\,.
\end{equation}
Thus, in the large distance limit we have 
\begin{equation}
    \langle j^{a}(x,z)j^{b}(x',z')\rangle=-\frac{\Gamma[\frac{d+1}{2}]}{(d-1)\Gamma[d]}\frac{(\Delta_{+}-\Delta_{-})^{2}\pi^{\frac{1-d}{2}}\Gamma[\Delta_{+}]\Gamma[\Delta_{-}]}{2\Gamma[\Delta_{+}-\frac{d-3}{2}]\Gamma[\Delta_{-}-\frac{d-3}{2}]}AB\partial^{a}\partial'^{b}G_{\text{massless}}(X,X')\,,
\end{equation}
and using Equ.~(\ref{eq:Cholographic}) we have
\begin{equation}
     \langle j^{a}(x,z)j^{b}(x',z')\rangle=-2g^{2}C_{1}C_{2}\frac{\pi^{\frac{d-1}{2}}\Gamma[\frac{d-1}{2}]}{\Gamma[d]}\partial^{a}\partial'^{b}G_{\text{massless}}(X,X')\,.
\end{equation}
We can get the dual boundary $C_{J}$ from the bulk gauge field two-point function following \cite{D'Hoker:1999jc} as
\begin{equation}
    C_{J}=\frac{(d-3)\Gamma[d-1]g_{e}^{2}}{2\pi^{\frac{d-1}{2}}\Gamma[\frac{d-1}{2}]}\,.
\end{equation}
As a result, we have
\begin{equation}
     \langle j^{a}(x,z)j^{b}(x',z')\rangle=-\frac{g^{2}}{g_{e}^{2}}\frac{C_{1}C_{2}}{C_{J}}\frac{d-3}{d-1}\partial^{a}\partial'^{b}G_{\text{massless}}(X,X')=-\frac{m^{2}_{\gamma}}{q^{2}g_{e}^{2}}\partial^{a}\partial'^{b}G_{\text{massless}}(X,X')\,,
\end{equation}
where we have $m_{\gamma}^2=g^2 q^2\frac{d-3}{d-1}\frac{C_1 C_2}{C_{J}}$ from Equ.~(\ref{eq:photonmass}). Now we are ready to evaluate the effective action Equ.~(\ref{eq:Seff})
\begin{equation}
\begin{split}
    S_{\text{eff},2}[\theta,0]&=i\frac{1}{2}\int d^{d-1}xdz d^{d-1}x'dz' \sqrt{-g}\sqrt{-g'}\partial_{a}\theta(x,z)\langle j^{a}(x,z)j^{b}(x',z')\rangle\partial'_{b}\theta(x',z')\,,\\&=-i\frac{m_{\gamma}^{2}}{q^{2}}\frac{1}{2}\int d^{d-1}xdz d^{d-1}x'dz' \sqrt{-g}\sqrt{-g'}\partial_{a}\theta(x,z)\partial^{a}\partial'^{b}G_{\text{massless}}(X,X')\partial'_{b}\theta(x',z')\,,\\&=i\frac{m_{\gamma}^{2}}{q^{2}}\frac{1}{2}\int d^{d-1}xdz d^{d-1}x'dz' \sqrt{-g}\sqrt{-g'}\theta(x,z)\partial_{a}\partial^{a}\partial'^{b}G_{\text{massless}}(X,X')\partial'_{b}\theta(x',z')\,,
    \label{eq:Seffm}
    \end{split}
\end{equation}
where in the last step we integrated by parts and used the fact that 
\begin{equation}
    G_{\text{massless}}(x,z)\sim z^{d}\,,\quad \theta(x,z)\rightarrow0\,, \quad\text{as }z\rightarrow0\,,
\end{equation}
to set the boundary piece zero.

This can be further simplified if one uses the equation of motion of the massless scalar field Greens function
\begin{equation}
\frac{1}{\sqrt{-g}}\partial_{a}\sqrt{-g}\partial^{a}G_{\text{massless}}(X,X')=i\frac{1}{\sqrt{-g}}\delta^{d}(X-X')\,,
\end{equation}
which results in
\begin{equation}
    S_{\text{eff,2}}[\theta,0]=-\frac{m_{\gamma}^{2}}{q^{2}}\frac{1}{2}\int d^{d-1}xdz \sqrt{-g}\partial_{a}\theta(x,z)\partial^{a}\theta(x,z)\,.
\end{equation}
Finally, we have the full quadratic effective induced action from integrating out the transparent matter
\begin{equation}
    S_{\text{eff,2}}[\theta,A]=-\frac{m_{\gamma}^{2}}{2}\int d^{d-1}xdz \sqrt{-g}\Big(A_{a}(x,z)-\frac{1}{q}\partial_{a}\theta(x,z)\Big)\Big(A^{a}(x,z)-\frac{1}{q}\partial^{a}\theta(x,z)\Big)\,,\label{eq:finalaction}
\end{equation}
which is obtained by replacing $\partial_{a}\theta$ by the invariant combination Equ.~(\ref{eq:invariantcomb}). The result Equ.~(\ref{eq:finalaction}) is exactly the photon mass term with mass square given by the field theory prediction Equ.~(\ref{eq:photonmass}) in the St\"{u}ckelberg form. Thus, we see that the photon in the gravitational AdS part of the island mode has a mass that corresponds to the U(1) gauge symmetry being spontaneously broken at low energy.
.
\subsubsection{The Gauge Symmetry is Mixed with the Global Symmetry}
The above results on symmetry breaking and the Higgs mechanism have a nice field theory dual. In this subsection, we will show that this field theory dual further suggests that the spontaneously broken gauge symmetry is mixed with the bath global symmetry.

At the level of the field theory analysis we performed in Sec.~\ref{sec:CFTanalysis}, this follows from Equ.~(\ref{eq:noether1}) which states that the current operator $J_{1}^{\mu}$ has a descendant $-igq\Big(O_{1}(x)O_{2}(x)-O_{1}^{*}(x)O_{2}^{*}(x)\Big)$ so the conformal multiplet associated with the current operator $J_{1}^{\mu}$ is no longer a short multiplet as for conserved currents but a long multiplet. In the AdS dual, this long multiplet becomes a long multiplet of the AdS isometry group,  implying that the vector boson has an extra longitudinal polarization that is dual to $\Big(O_{1}(x)O_{2}(x)-O_{1}^{*}(x)O_{2}^{*}(x)\Big)$. This is exactly the Higgs mechanism where the vector boson eats a scalar field, that becomes its longitudinal polarization mode, and becomes massive. This extra polarization mode is the Goldstone boson associated with the spontaneous breaking of the U(1) gauge symmetry. Let's work out this observation in some detail below.

First, we note that the bath coupling will induce a vacuum expectation value for the operators
\begin{equation}
    O_{1}(x)O_{2}(x)\,,\quad O_{1}^{*}(x)O^{*}_{2}(x)\,,
\end{equation}
as to leading order in the coupling $g$ we have
\begin{equation}
    \langle  O_{1}(x)O_{2}(x)\rangle=\langle  O^{*}_{1}(x)O^{*}_{2}(x)\rangle=g\int d^{d-1}y\frac{C_{1}C_{2}}{|x-y|^{2(d-1)}}\equiv v\,,
\end{equation}
where we used the fact that $\Delta_{1}+\Delta_{2}=d-1$ for marginal couplings and the result is UV divergent due to the lack of a scale. Introducing a a UV cutoff length scale $\epsilon$ we have
\begin{equation}
    v\sim \frac{1}{\epsilon^{d-1}}\,.\label{eq:vepsilon}
\end{equation}
At finite energy, the phase of $O_{1}(x)O_{2}(x)$ and $ O^{*}_{1}(x)O^{*}_{2}(x)$ is a dynamical field that decouples from other sectors as the couplings of the this phase and other sectors are suppressed by $\frac{1}{v}$. Let's denote this phase as $\Phi(x)$, i.e.
\begin{equation}
    O_{1}(x)O_{2}(x)=v e^{i\Phi(x)}\,,
\end{equation}
then $\Phi(x)$ transforms non-linearly under the two original U(1) symmetries generated by the CFT$_{d-1}$ and bath currents $J_{1}^{\mu}$ and $J_{2}^{a}$ when the AdS and bath are decoupled but is neutral under the global symmetry generated by Equ.~(\ref{eq:globalJ}).

The holographic dual of $\Phi(x)$ transforms non-linearly under the gauge symmetry U(1). We will denote this holographic dual as $\theta(x,z)$ where $z$ is the radial coordinate in the AdS$_{d}$ Poincar\'{e} patch. By holographic duality, $\theta(x,z)$ transforms nonlinearly under the bulk gauge symmetry whose constant part is dual to the global symmetry generated $J_{1}^{\mu}$. $\Phi(x)$ also transforms nonlinearly under the bath global symmetry generated by $J_{2}^{\mu}$. Thus, a certain linear combination of the global symmetries generated by $J_{1}^{\mu}$ and $J_{2}^{\mu}$ leaves $\Phi(x)$ invariant. This linear combination will also leave $\theta(x,z)$ invariant by holographic duality. This implies that $\theta(x,z)$ also transforms nonlinearly under the global symmetry generated by $J_{2}^{\mu}$.  With the canonical normalization of $\Phi(x)$, i.e. $\tilde{\Phi}(x)=v\Phi(x)$, the phase decouples from other sectors and ensures that
\begin{equation}
    \partial_{\mu} J^{\mu}_{1}(x)=gq\tilde{\Phi}(x)+\mathcal{O}(\frac{1}{v})\neq0\,,\label{eq:nonconservedPhi}
\end{equation}
so one can think that $J_{1}^{\mu}(x)$ eats an  extra mode $\tilde{\Phi}(x)$ which is decoupled from other sectors, i.e. an operator Higgs mechanism. From Equ.~(\ref{eq:nonconservedPhi}), we  see that the operator $\tilde{\Phi}(x)$ has conformal dimension $\Delta_{\tilde{\Phi}}=d-1$ to leading order in the coupling. This suggests that the holographic relationship between $\Phi(x)$ and the bulk eaten Goldstone boson, $\theta(x,z)$ nearby the boundary is
\begin{equation}
    \theta(x,z)\sim z^{d-1}\tilde{\Phi}(x)\,,\quad z\rightarrow0\,,\label{eq:thetaholography}
\end{equation}
where $z$ is the radial coordinate in the AdS$_{d}$ Poincar\'{e} patch. 

The above result Equ.~(\ref{eq:thetaholography}) is consistent with the fact that $\theta(x,z)$ is a massless field.\footnote{The dual operator of $\theta(x,z)$ has conformal weight $\frac{d-1}{2}+\sqrt{\frac{(d-1)^{2}}{4}+m^{2}}=d-1$ which is consistent with $\Delta_{\Phi_{n}}=d-1$.} Moreover, Equ.~(\ref{eq:thetaholography}) is consistent with the fact that $\theta(x,z)$ transforms non-linearly under the coupling induced broken global symmetry generated by $J^{\mu}_{1}(x)$. This is because, from Equ.~(\ref{eq:thetaholography}), we can see that under the above U(1) transform in the bulk with the transformation parameter $\alpha$ we have
\begin{equation}
    \tilde{\Phi}(x)\rightarrow\tilde{\Phi}(x)+\frac{q}{z^{d-1}}\alpha\,,\quad z\rightarrow0\,,
\end{equation}
which is consistent with what one would infer if one instead studies the U(1) transform of $\Phi(x)$ directly on the field theory side, i.e.
\begin{equation}
    \tilde{\Phi}(x)\rightarrow\tilde{\Phi}(x)+q v\alpha=\tilde{\Phi}(x)+\frac{q}{\epsilon^{d-1}}\alpha\,,
\end{equation}
where we have used Equ.~(\ref{eq:vepsilon}).

In summary, we see that in the dual island setup, we have an exact global U(1) symmetry generated Equ.~(\ref{eq:globalJ}) which is mixed with an AdS spontaneously broken gauge symmetry $U(1)_{\text{gauge}}$ that is holographically dual to $J_{1}^{\mu}$. The charges of various operators under these two symmetry transformations and the two coupling induced broken global symmetry transformations are listed in Table.\ref{tab:charge}, where the charges are to zeroth order in the coupling $g$. To understand the column for $U(1)_{\text{global}}$ in Table.\ref{tab:charge} we note that from Equ.~(\ref{eq:globalJ}), we have its charge operator as $Q_{\text{global}}=q'Q_{\text{global broken, 1}}-qQ_{\text{global broken, 2}}$.  The first two rows say the global charge for an operator $a(x)$ in the AdS that had gauge charge $q_a$  and for an operator $b(x)$ in the bath that had charge $q_b$ under the original global symmetry of the bath. The exact global symmetry and the spontaneously broken gauge symmetry are mixed in the sense that the AdS scalar field $\phi_{1}(x,z)$ is charged under both of them.

\begin{table}
\centering
\caption{The charges of various operators.}
\begin{tabular}{|p{4em}|p{3cm}|p{4em}|p{7em}|p{7em}|}
\hline
\text{ }& U(1)$_{\text{global}}$ & U(1)$_{\text{gauge}}$ & U(1)$_{\text{global broken,1}}$ &U(1)$_{\text{global broken,2}}$\\
 \hline
 $a(x)$ &$q'q_{a}$&$q_{a}$&$q_a$&$0$\\\hline
 $b(x)$&$-qq_{b}$&$0$&$0$&$q_{b}$\\\hline
$O_{1}(x)$  & $qq'$  & 0 &$q$&0 \\
\hline
$O_{2}(x)$  & $-qq'$  & 0&0&$q'$ \\
\hline
$\Phi(x)$  & 0  & 0&$q_{\text{non-linear}}$&$q'_{\text{non-linear}}$ \\
\hline
$\phi_{1}(x,z)$  & $qq'$  & $q$&$q$& 0 \\
\hline
$\theta(x,z)$  & 0  & $q_{\text{non-linear}}$&$q_{\text{non-linear}}$& $q'_{\text{non-linear}}$\\
\hline
\end{tabular}\label{tab:charge}
\end{table}

\subsubsection{Operators in the Island and Their Global Symmetry Charges}\label{sec:constraint}
Having understood the fate of the symmetries in the island setup, we are ready to study charged operators. For this purpose, let's consider a complex scalar single-trace primary operator $\hat{O}_{q_{1}}(x)$ in the CFT$_{d-1}$ which is charged under the coupling-induced broken global U(1) symmetry generated by $J_{1}^{\mu}$. Unlike the operator $\hat{O}_{1}(x)$, this operator is not directly coupled to the bath. The AdS bulk dual of this charged operator is a complex free massive scalar field $\phi(x,z)$ with charge $q_{1}$ under the U(1) gauge symmetry and obeying the reflective boundary condition near the asymptotic boundary. Both the CFT operator $\hat{O}_{q_{1}}(x)$ and its bulk dual $\phi(x,z)$ have charge $q_{1}q'$  under the unbroken global symmetry generated by Equ.~(\ref{eq:globalJ}) and charge zero under the coupling-induced broken bath global symmetry generated by $J_{2}^{a}$.

The relevant action in the AdS is
\begin{equation}
\begin{split}
    S_{\text{AdS, eff}}=&-\int d^{d-1}xdz\sqrt{-g}\Big[g^{ab}D_{a}\phi D^{*}_{b}\phi^{*}+M^{2}\phi\phi^{*}\Big]-\frac{1}{4g_{e}^{2}}\int d^{d-1}x dz\sqrt{-g} F_{ab}F^{ab}\\&-\frac{m_{\gamma}^{2}}{2}\int d^{d-1}xdz \sqrt{-g}\Big(A_{a}-\frac{1}{q}\partial_{a}\theta\Big)\Big(A^{a}-\frac{1}{q}\partial^{a}\theta\Big)\,,\label{eq:actionAdSR}
    \end{split}
\end{equation}
where the gauged covariant derivative is now $D_{a}=\partial_{a}-iq_{1} A_{a}(x,z)$. Operators and states in the AdS have to obey the Gauss' law constraint
\begin{equation}
    iq_{1}g^{0b}\phi D_{b}^{*}\phi^{*}-iq_{1}g^{0b}D_{b}\phi \phi^{*}+\frac{1}{g_{e}^{2}}\partial_{a}F^{a0}-\frac{m_{\gamma}^{2}}{q} g^{0b}(A_{b}-\partial_{b}\theta)=0\,.\label{eq:gauss}
\end{equation}
Since we are  interested only in local operators, we can equivalently study the integrated constraint which is the integral of the Equ.~(\ref{eq:gauss}) along a spatial slice that contains the operator we are interested in. Before we write down the integrated constraint, we introduce the canonical momenta
\begin{equation}
\begin{split}
    \Pi_{\phi}(t)&=\int d^{d-2}\vec{x}dz\sqrt{-g}g^{0b} D_{b}^{*}\phi^{*}\,,\quad \Pi_{\phi^{*}}(t)=\int d^{d-2}\vec{x}dz\sqrt{-g}g^{0b} D_{b}\phi\,,\\\Pi_{A_{i}}(t)&=-\int d^{d-2}\vec{x}dz\sqrt{-g}\frac{1}{g_{e}^{2}}F^{0i}\,,\quad \Pi_{\theta}(t)=\frac{m_{\gamma}^{2}}{q^{2}}\int d^{d-2}\vec{x}dz\sqrt{-g}g^{0b}(A_{b}-\partial_{b}\theta)\,,
    \end{split}
\end{equation}
where we used $i$ to denote spatial directions and the integrations are all along a spatial slice. We have canonical equal-time commutators,
\begin{equation}
    \begin{split}
        [ \Pi_{\phi}(t),\phi(x,t)]=i\,,\quad [ \Pi_{\phi^{*}}(t),\phi^{*}(x,t)]=i\,,\quad[\Pi_{A_{i}}(t),A_{j}(t)]=i\delta_{ij}\,,\quad [\Pi_{\theta}(t),\theta(t)]=i\,.
    \end{split}
\end{equation}
The integrated constraint of Equ.~(\ref{eq:gauss}) can be written as
\begin{equation}
  iq_{1} \int d^{d-2}\vec{x}dz \sqrt{-g}\Big(\phi\pi_{\phi}-\phi^{*}\pi_{\phi^{*}}\Big)+\int_{z=0} d^{d-2}\vec{x} \sqrt{-g} \pi_{A_{z}}-q\Pi_{\theta}=0\,,
\end{equation}
where we used $\pi$ to denote the canonical momentum densities. The first term is  the charge operator which we will denote $Q_{\text{matter}}$. So have the integrated constraint
\begin{equation}
    Q_{\text{matter}}+\int_{z=0} d^{d-2}\vec{x} \sqrt{-g} \pi_{A_{z}}-q\Pi_{\theta}=0\,.\label{eq:constraint}
\end{equation}
When this constraint is imposed on operators, we have to ensure that the constrained operator commutes with the operator on the left-hand side of Equ.~(\ref{eq:constraint}), which simply means that the operator has to be invariant under small gauge transforms. We will take the field $\hat{\phi}(x,z)$ itself as an explicit example. This operator is charged under the U(1) gauge symmetry so we should dress it to obey the following constraint
\begin{equation}
    [\hat{Q}_{\text{matter}}+\int_{z=0} d^{d-2}\vec{x} \sqrt{-g} \hat{\pi}_{A_{z}}-q\hat{\Pi}_{\theta},\hat{\phi}^{\text{Phys}}(x,z)]=0\,.\label{eq:Oconstraint}
\end{equation}
Due to the spontaneous symmetry breaking, there are two ways to dress $\hat{\phi}(x,z)$ to a physical operator $\hat{\phi}^{\text{Phys}}(x,z)$ which obeys the above constraint. The first way is the standard Wilson line dressing fron the bulk point $(x,z)$ to the boundary $z=0$ which gives
\begin{equation}
    \hat{\phi}^{\text{Phys}}_{\bar{I}}(x,z)=e^{iq_{1}\int_{\mathcal{C}}\hat{A}_{a}dx^{a}}\hat{\phi}(x,z)\,,\label{eq:outisland}
\end{equation}
where $\mathcal{C}$ is a spatial contour that connects the bulk point $(x,z)$ to the boundary $z=0$.

But there is an alternative way to dress the operator that is key to understanding the island charge. This second way is to dress using  $\theta(x,z)$
\begin{equation}
    \hat{\phi}^{\text{Phys}}_{I}(x,z)=e^{-i\frac{q_{1}}{q}\hat{\theta}(x,z)}\hat{\phi}(x,z)\,.\label{eq:inisland}
\end{equation}
It is easy to check that the above operators Equ.~(\ref{eq:outisland}) and Equ.~(\ref{eq:inisland}) obey the constraint Equ.~(\ref{eq:Oconstraint}).\footnote{We note that there is in fact a third way the dress operators to obey the constraint Equ.~(\ref{eq:Oconstraint}). This works only for special cases where we have oppositely charged operators \cite{Francois:2024rdm}. In this case, one can dress the oppositely charged operators using a Wilson line connecting them. We will not consider this possibility as the total global symmetry charge for the operators of this type is zero.} though the ways they obey the constraint are rather different. Equ.~(\ref{eq:outisland}) obeys the constraint by dressing the original charged operator $\hat{\phi}(x,z)$ from $(x,z)$ to the boundary using a line operator and this dressing creates an operator from a  charged operator situated at $(x,z)$ to a an operator with boundary $z=0$ charge.  Hence the dressed operator Equ.~(\ref{eq:outisland}) can be detected near the boundary by measuring its correlation functions with the near boundary gauge fields. This is the standard dressing in gauge theories. 

However, as opposed to the line-dressed operator Equ.~(\ref{eq:outisland}), Equ.~(\ref{eq:inisland}) is locally dressed using the local operator $\theta(x,z)$\footnote{We should emphasize $\theta(x,z)$ is local from the perspective inside the AdS and as we have discussed in Sec.~\ref{sec:CFTanalysis} $\theta(x,z)$  doesn't commute with a bath operator.} so the operator Equ.~(\ref{eq:inisland}) cannot be detected near the asymptotic boundary of the AdS by measuring the electric field.

As was pointed out in \cite{Geng:2020qvw,Geng:2021hlu,Geng:2023zhq}, a basic consistency condition for the holographic interpretation of entanglement islands is that the gravitational AdS operators inside the island have to commute with operators outside the island. This consistency condition enforces the second way of dressing for operators inside the island $\mathcal{I}$. Moreover, \cite{Geng:2025rov} pointed out another basic consistency condition for the holographic interpretation of entanglement islands that operators inside the island have to be detectable by operators in the bath. For example, we can apply an operator in the island to create a local energetic excitation inside the island. Given the duality, the bath Hamiltonian should not commute with operators in the island. 

In our case, a similar condition should hold. 
This is true as the dressed operator is charged under the unbroken U(1) global symmetry. 
We can see this by observing the operator Equ.~(\ref{eq:inisland}) is neutral under the coupling induced broken global symmetry generated by $J_{1}^{\mu}$ of the dual CFT$_{d-1}$ as this symmetry is dual to the bulk gauge symmetry under which the operator Equ.~(\ref{eq:inisland}) by construction is neutral. The operator dressed to the bath has  charge $q_{1}q'$  under the exact global U(1) symmetry generated by Equ.~(\ref{eq:globalJ}) as the original AdS field $\phi(x,z)$ has charge $q_{1}q'$  and the Goldstone boson $\theta(x,z)$ has zero charge under this exact global symmetry (see Table.\ref{tab:charge}). Since the global symmetry under which the $\theta$-dressed operator in AdS is generated by a charge supported in both the CFT$_{d-1}$ and in the bath and this operator is neutral for the CFT$_{d-1}$ part, the global charge of this operator is encoded  in the bath. This consistency condition follows since the Goldstone boson $\theta(x,z)$ is indeed charged under the coupling-induced broken bath global U(1) symmetry generated by $J_{2}^{a}$.  As a result, the operator  Equ.~(\ref{eq:inisland}) can  be detected by the bath operator\footnote{In fact to ensure that $Q_{\text{bath}}$ is well defined one has to properly smear the current $J_{2}^{a}$ using a smearing kernel which goes to zero smoothly near the boundary. We stay to the lowest order in perturbation theory so we will ignore this issue.}
\begin{equation}
    Q_{\text{bath}}=\int d^{d-1}\vec{x} J_{2}^{0}(x)\,.
\end{equation}
See Table.\ref{tab:charge2} for the charges under various symmetries. 

This is particularly significant inside the island where We see that the global symmetry charges of operators in the island are fully encoded in the bath and are invisible to the CFT$_{d-1}$. This is consistent with our analysis in Sec.~\ref{sec:general} that global symmetries are consistent with the entanglement island, i.e. the black hole interior, as global symmetry charges inside the island are invisible to the dual CFT$_{d-1}$, i.e. the quantum mechanical system that describes the black hole when the bath is decoupled.

\subsection{Summary and Discussions}
In summary, in this section we performed an in-depth study of symmetries in the island setup. We found that with leaky charged matter in the gravitational AdS part of the island setup, the gauge symmetry is spontaneously broken and is mixed with an exact global symmetry. The study of the gauge symmetry and its spontaneous breaking was performed in  both the dual field description of the island setup and the island setup itself with various exactly matched results. This study extends the results and proved conjectures in various old and recent papers \cite{Porrati:2001gx,Porrati:2002dt,Porrati:2003sa,Duff:2004wh,Aharony:2006hz,Rattazzi:2009ux,Karch:2023wui} with an explicit demonstration of the existence of a Goldstone boson, which is eaten by the gauge field in the Higgs mechanism, and the understanding of its field theory dual. The Goldstone boson is non-locally supported in the union of the AdS and the bath, though an observer inside the AdS will observe it as a local field. This fact enables us to identify operators inside the island that are consistent with the holographic interpretation of the island. Moreover, the Goldstone boson is non-linearly charged under various coupling-induced broken global symmetries including the one of the bath and is neutral under the exact global symmetry. This property ensures that the global symmetry charges of operators inside the island will not be changed due to the dressing and can be detected from the bath but are invisible to the boundary CFT$_{d-1}$. This is consistent with the mechanism uncovered in \cite{Geng:2025rov} for the information in the island being fully encoded in the bath. This mechanism is fully consistent with our analysis in Sec.~\ref{sec:general} regarding the quantum mechanics of black holes and global symmetries. Moreover, our construction and analysis in this section can be easily generalized to non-Abelian gauge symmetries.

\begin{table}
\centering
\caption{The charges of various AdS operators.}
\begin{tabular}{|p{4.5em}|p{3cm}|p{4em}|p{7em}|p{7em}|}
\hline
\text{ }& U(1)$_{\text{global}}$ & U(1)$_{\text{gauge}}$ & U(1)$_{\text{global broken,1}}$ &U(1)$_{\text{global broken,2}}$\\
 \hline
$\phi(x,z)$  & $q_{1}q'$  & $q_{1}$&$q_{1}$& 0 \\
\hline
$\theta(x,z)$  & 0  & $q_{\text{non-linear}}$&$q_{\text{non-linear}}$& $q'_{\text{non-linear}}$\\
\hline
$\phi^{\text{Phys}}(x,z)$  & $q_{1}q'$  & 0&0& $-\frac{q'q_{1}}{q}$\\
\hline
\end{tabular}\label{tab:charge2}
\end{table}

\section{The Higher-Dimensional Holographic Dual-- Karch-Randall Braneworld}\label{sec:KR}
    In Sec.~\ref{sec:GlobalSymIsland}, we studied a model with an island in which the AdS and the bath are assumed to be weakly coupled with a double-trace deformation Equ.~(\ref{eq:CFTdescription}). We uncovered various important aspects of this model, including the existence of global symmetries and the mechanism for the global symmetry charges of operators inside the island to be fully encoded in the bath. We studied both the island theory directly and its field theory dual. Nevertheless, it was not very intuitive how the global symmetry charges inside the island are encoded in the bath. In this section, we will study the  Karch-Randall braneworld \cite{Karch:2000ct,Karch:2000gx,Geng:2023qwm}, which has a strongly coupled holographic dual island setup. We will see how the mechanism we uncovered in Sec.~\ref{sec:CFTanalysis} is nicely geometrized to an intuitive picture in this framework. This picture will be helpful in understanding the resolution of the Harlow-Shaghoulian paradox in Sec.~\ref{sec:resolvingHS}.
 
\subsection{The Karch-Randall Braneworld}
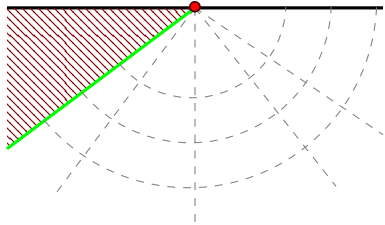
\begin{figure}[h] 
\begin{centering}
\begin{tikzpicture}[scale=1]
\draw[-,very thick,black!100] (-2.5,0) to (0,0);
\draw[-,very thick,black!100] (0,0) to (2.5,0);
\draw[pattern=north west lines,pattern color=purple!200,draw=none] (0,0) to (-2.5,-1.875) to (-2.5,0) to (0,0);
\draw[-,dashed,color=black!50] (0,0) to (-1.875,-2.5); 
\draw[-,dashed,color=black!50] (0,0) to (0,-2.875);
\draw[-,dashed,color=black!50] (0,0) to (1.875,-2.375); 
\draw[-,dashed,color=black!50] (0,0) to (2.5,-1.6875); 
\draw[-,very thick,color=green!!50] (0,0) to (-2.5,-1.875);
\node at (0,0) {\textcolor{red}{$\bullet$}};
\node at (0,0) {\textcolor{black}{$\circ$}};
\draw[-,dashed,color=black!50] (-2,-1.5) arc (-140:-2:2.5);
\draw[-,dashed,color=black!50] (-1.5,-1.125) arc (-140:-2:1.875);
\draw[-,dashed,color=black!50] (-1,-0.75) arc (-140:-2:1.25);
\end{tikzpicture}
\caption{\small A constant time slice of an AdS$_{d+1}$ with a Karch-Randall brane. The green surface denotes the brane and it has AdS$_{d}$ geometry. All the dashed straight lines are in fact legitimate places for the brane to reside with the specific slice selected according to the brane tension \cite{Geng:2023qwm}. The purple-shaded region behind the brane is cutoff. The red dot is the codimension one submanifold of the bulk asymptotic boundary (the thick black line) where it intersects the brane. This submanifold is also called the \textit{defect} in the literature. }
\label{pic:branedemo}
\end{centering}
\end{figure}

In this subsection, we provide a lightning review of the Karch-Randall braneworld, emphasizing its relevance to our study of models with islands. The Karch-Randall braneworld concerns the physics of an AdS$_{d}$ brane embedded in an ambient gravitational AdS$_{d+1}$ space. The brane is glued to the asymptotic boundary of the bulk AdS$_{d+1}$ along a codimension one submanifold (see Fig.\ref{pic:branedemo}). This codimension one submanifold of the bulk asymptotic boundary is also the asymptotic boundary of the AdS$_{d}$ brane. The brane cuts off the part of the bulk region behind it, but nonetheless leaves the other half of the boundary. This gives a $d$-dimensional gravitational theory but one with a massive graviton.

One can also deduce the existence of the gravitational theory by applying the AdS/CFT correspondence to the bulk to dualize it to a gravitational theory on the brane coupled to a conformal field theory living on the leftover asymptotic boundary through a transparent boundary condition. This is consistent with  the fact that the graviton on the brane has a mass, which  can be seen by doing a KK analysis of the bulk graviton reduced from AdS$_{d+1}$ to AdS$_{d}$ \cite{Karch:2000ct,Geng:2025rov,Geng:2025tba3}. We have already seen this result in the description of the theory with an island-- the intermediate description-- in which the AdS$_{d}$ graviton becomes massive due to bath coupling  that induces spontaneous  breaking of the diffeomorphism symmetry in the AdS$_{d}$ \cite{Porrati:2001gx,Porrati:2002dt,Porrati:2003sa,Duff:2004wh,Aharony:2006hz,Geng:2023qwm,Geng:2023ynk,Geng:2025rov}, analogously to the discussion of Sec.~\ref{sec:GlobalSymIsland} for the gauge field. Thus the Karch-Randall braneworld is a holographic model for the island setup, where however the matter fields are strongly coupled as for any holographic theory. 
The dual island setup is often referred to as the \textit{intermediate description} of the Karch-Randall braneworld \cite{Geng:2020qvw,Geng:2020fxl,Geng:2023qwm}. 

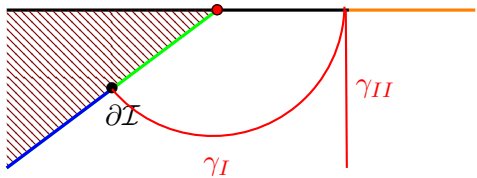
\begin{figure}[h]
\begin{centering}
\begin{tikzpicture}[scale=1.4]
\draw[-,very thick,black!100] (-2,0) to (0,0);
\draw[-,very thick,black!100] (0,0) to (1.25,0);
\draw[-,very thick,orange] (1.25,0) to (2.45,0);
\draw[pattern=north west lines,pattern color=purple!200,draw=none] (0,0) to (-2,-1.5) to (-2,0) to (0,0);
\draw[-,very thick,color=green!!50] (0,0) to (-2,-1.5);
\node at (0,0) {\textcolor{red}{$\bullet$}};
\node at (0,0) {\textcolor{black}{$\circ$}};
\draw[-,very thick,blue] (-1,-0.75) to (-2,-1.5);
\node at (-1,-0.75) {\textcolor{black}{$\bullet$}};
\node at (-0.9,-1) {\textcolor{black}{$\partial\mathcal{I}$}};
\draw[-,thick,color=red] (-1,-0.75) arc (-140:-1:1.25);
\draw[-,thick,color=red] (1.22,0) to (1.23,-1.5);
\node at (0,-1.49) {\textcolor{red}{$\gamma_{I}$}};
\node at (1.5,-0.75) {\textcolor{red}{$\gamma_{II}$}};
\end{tikzpicture}
\caption{\small A demonstration of the construction of entanglement island in the Karch-Randall braneworld. The orange line denotes the bath subregion $R$ with blue region on the brane denoting its entanglement island $\mathcal{I}$. The red surface $\gamma_{I}$ connecting $\partial R$ and $\partial\mathcal{I}$ is a minimal area surface. $\gamma_{I}$ is determined by minimizing its area also with respect to its ending point $\partial \mathcal{I}$ on the brane which correspondence to the $\min_{ \mathcal{I}}$ in the island formula Equ.~(\ref{eq:islandformula}). $\gamma_{I}$ and $\gamma_{II}$ are called the Ryu-Takayanagi surfaces.}
\label{pic: branwisland}
\end{centering}
\end{figure}

Consistent with the above interpretation,  there are  entanglement islands that can be found using holographic tools in this setup. The entanglement entropy of a bath subregion $R$ can be computed using the modified Ryu-Takayanagi (RT) formula \cite{Ryu:2006bv,Ryu:2006ef,Fujita:2011fp,Takayanagi:2011zk,Miao:2017gyt}. As opposed to the RT formula in the standard AdS/CFT setup, the RT surface can now end on the Karch-Randall brane and one also has to minimize the surface area of the RT surface over its possible ending point on the brane \cite{Chen:2020uac,Chen:2020hmv,Geng:2020qvw,Geng:2020fxl}. This minimization over the brane ending point is actually the holographic dual of the minimization over possible island regions in the island formula Equ.~(\ref{eq:islandformula}). This minimization prescription is derived using replica wormholes \cite{Geng:2024xpj} following the proof of the RT formula in the standard AdS/CFT correspondence \cite{Lewkowycz:2013nqa,Dong:2013qoa,Camps:2013zua,Miao:2014nxa}. Thus we have
\begin{equation}
    S(R)=\min\Big(\frac{A(\gamma_{I})}{4G_{N}},\frac{A(\gamma_{II})}{4G_{N}}\Big)\,,\label{eq:RTisland}
\end{equation}
where $G_{N}$ is the AdS$_{d+1}$ bulk Newton's constant, $\gamma_{I}$ is the minimal area surface of the type that is not ending on the brane and $\gamma_{II}$ is the minimal area surface that ends on the brane and has been minimized over the possible ending point on the brane (see Fig.\ref{pic: branwisland} for a demonstration). For a choice of the bath subregion $R$ whose boundary is close enough to the defect, the second type of  RT surface dominates and the output of the formula Equ.~(\ref{eq:RTisland}) is the entanglement entropy of $R$ along with its entanglement island $\mathcal{I}$ on the brane. In this setup, it is more intuitive to see that the entanglement island $\mathcal{I}$ lives in the entanglement wedge of the bath subregion $R$ as it follows from the definition of entanglement wedge in the standard AdS/CFT setup applied to the full higher dimensional geometry.

\subsection{Global and Gauge Symmetries in the Karch-Randall Braneworld}\label{sec:globalgaugeKR}
. 

We now consider a dynamical gauge field living in the AdS$_{d+1}$ bulk and we will show that the holographic dual intermediate description is analogous to the model studied in Sec.~\ref{sec:GlobalSymIsland} with a {spontaneously} broken gauge symmetry and a preserved global symmetry.  So far, we have considered the CFT dual of the island setup as well as the island setup itself. In this section, we employ double holography to consider the Karch-Randall bulk dual description of  the brane island setup. 

The holographic dual intermediate description can be decoded from the $(d+1)$-dimensional bulk by Kaluza-Klein (KK) decomposition. Thus, we will perform the KK decompositions for the bulk gauge field and the bulk gauge transform. The result is a tower of massive $d$-dimensional gauge fields and the corresponding gauge transforms. These gauge fields are the gauge fields in the holographic dual island setup. We will see that  (as with the graviton \cite{Karch:2000ct}) there is no normalizable zero mode.\footnote{In a future paper \cite{Geng:2026tba1}, we will argue that as with the graviton, there is an anomalously light mode that reduces to the zero mode in the zero curvature limit.} Thus, all the $d$-dimensional gauge fields are massive.\footnote{We note that this extra-dimensional Wilson line is distinct from the Wilson line along the brane directions as it looks like a local field from the perspective of the brane, rather than a nonlocal line operator.}

Now we perform the explicit analysis. The upshot of these calculations will be that in the KR picture the dressing of operators via $\theta$ can be realized by a Wilson line traversing the bulk spacetime as depicted in Fig.\ref{pic:branethetadress}. Part of this system has been studied in \cite{Geng:2025rov}, with which our discussion in the following overlaps. We will write the most general (d+1)-dimensional warped geometry in the bulk before we restrict to AdS$_{d+1}$. We consider the bulk metric
\begin{equation}
ds^2=d\rho^{2}+e^{2A(\rho)}\bar{g}_{ij}(x)dx^{i}dx^{j}\,,\label{eq:metric}
\end{equation}
When the bulk is AdS$_{d+1}$, the brane position $\rho_{B}$ is related to the brane tension $T$ as $T=(d-1)\tanh\rho_{B}$ \cite{Geng:2023qwm} with the bulk $\rho$-coordinate $\rho\in[\rho_{B},\infty)$.

We study a U(1) gauge field in this bulk geometry.    Since we have a free theory, we can deduce the action from the equation of motion. We first look at the $j$-components of the (d+1)-dimensional equation of motion
\begin{equation}
\begin{split}
\nabla_{\mu}\mathbf{F}^{\mu}_{j}=e^{-2A(r)}\bar{\nabla}_{i}\mathbf{\bar{F}}^{i}_{j}+\Big[\partial^{2}_{\rho}+(d-2)A'(\rho)\partial_{\rho}\Big]\mathbf{A}_{j}-\Big[\partial_{j}\partial_{\rho}+(d-2)A'(\rho)\partial_{j}\Big]\mathbf{A}_{\rho}\,=0,
\end{split}
\end{equation}
where we defined $\bar{\mathbf{F}}^{i}_{j}=\bar{g}^{ik}\mathbf{F}_{kj}$. This equation is equivalent to the following equation
\begin{equation}
\bar{\nabla}_{i}\mathbf{F}^{i}_{j}=-e^{2A(\rho)}\Big[\partial^{2}_{\rho}+(d-2)A'(\rho)\partial_{\rho}\Big]\mathbf{A}_{j}+e^{2A(\rho)}\Big[\partial_{\rho}+(d-2)A'(\rho)\Big]\partial_{j}\mathbf{A}_{\rho}\,.\label{eq:dreduce}
\end{equation}
Furthermore, let's define a field $\theta(x,\rho)$ as follows
\begin{equation}
    \mathbf{\theta}(x,\rho)=-\int_{\mathcal{C}} \mathbf{A}\,,\label{eq:wilson}
\end{equation}
where $\mathcal{C}$ is contour that goes from $(x,\rho)$ to the asymptotic boundary along the $\rho$-direction. 
Thus, we have
\begin{equation}
    \mathbf{A}_{\rho}(x,\rho)=\partial_{\rho}\mathbf{\theta}(x,\rho)\,.
\end{equation}
Using this new field $\theta(x,\rho)$, we can write Equ.~(\ref{eq:dreduce}) as
\begin{equation}
\bar{\nabla}_{i}\bar{\mathbf{F}}^{i}_{j}=-e^{2A(\rho)}\Big[\partial^{2}_{\rho}+(d-2)A'(\rho)\partial_{\rho}\Big](\mathbf{A}_{j}-\partial_{j}\mathbf{\theta})\,.\label{eq:dreducetheta}
\end{equation}
The bulk gauge field $\mathbf{A}_{\mu}$ also has to satisfy the following boundary conditions close to the brane $\rho=\rho_{B}$
\begin{equation}
    \partial_{\rho}\mathbf{A}_{j}(x,\rho)|_{\rho=\rho_{B}}=0\,,\quad \mathbf{A}_{\rho}(x,\rho_{B})=0\,,\label{eq:maxwellbc}
\end{equation}
which ensures that the AdS$_{d}$ components $\mathbf{A}_{j}(x,\rho_{B})$ of the bulk gauge field as well as $\theta(x,\rho_{B})$ are free to fluctuate near the brane. This fluctuation near the brane is a signature of the existence of dynamical gauge fields on the AdS part in the dual island setup. With this boundary condition Equ.~(\ref{eq:maxwellbc}) imposed on the bulk gauge field $A_{\mu}(x,\rho)$, we can KK decompose its $j$-components and the Wilson line Equ.~(\ref{eq:wilson}) as
\begin{equation}
    \mathbf{A}_{j}(x,\rho)=\sum_{n=1}^{\infty}\psi_{n}(\rho)\bar{\mathbf{A}}^{(n)}_{j}(x)\,,\quad \theta(x,\rho)=\sum_{n=1}^{\infty}\theta^{(n)}(x)\psi_{n}(\rho)
\end{equation}
where $\bar{\mathbf{A}}^{(n)}_{j}(x)$ is the $n$-th KK mode and $\psi_{n}(\rho)$ is the wavefunction of the $n$-th KK mode that satisfies
\begin{equation}
    -e^{2A(\rho)}\Big[\partial^{2}_{\rho}+(d-2)A'(\rho)\partial_{\rho}\Big]\psi_{n}(\rho)=m_{n}^{2}\psi_{n}(\rho)\,,\quad\partial_{\rho}\psi_{n}(\rho)|_{\rho=\rho_{B}}=0\,,\label{eq:KKwave}
\end{equation}
with $m_{n}^{2}$ the mass square of the $n$-th KK mode. The normalization condition for the KK wavefunctions is 
\begin{equation}
    \int_{\rho_{B}}^{\infty} d\rho e^{(d-4)A(\rho)}\psi_{n}(\rho)\psi_{m}(\rho)=\delta_{mn}\,.\label{eq:normalization}
\end{equation}
As a result, we can see that if we project Equ.~(\ref{eq:dreducetheta}) to the normalizable eigenmodes $\psi_{n}(\rho)$ of the differential operator $-e^{2A(\rho)}\Big[\partial^{2}_{\rho}+(d-2)A'(\rho)\partial_{\rho}\Big]$, we get
\begin{equation}
    \bar{\nabla}_{i}\bar{\mathbf{F}}^{(n)i}_{j}(x)=m_{n}^{2}(\bar{\mathbf{A}}^{(n)}_{j}-\partial_{j}\mathbf{\theta}^{(n)})(x)\,,\label{eq:dreducethetan}
\end{equation}
which is exactly the equation of motion of a $d$-dimensional massive gauge field with the mass term in the St\"{u}ckelberg form. Hence we expect that in the dual island setup, we have a tower of decoupled massive gauge theories with St\"{u}ckelberg mass terms on the AdS part. As a check of this expectation, we can check if the equations of motion of the St\"{u}ckelberg fields $\theta^{(n)}(x)$ are satisfied. For this purpose, let's consider the $\rho$-th component of the bulk gauge field equation of motion
\begin{equation}
    \begin{split}
\nabla_{\mu}\mathbf{F}^{\mu}_{\rho}&=e^{-2A(\rho)}\bar{g}^{ij}\Big[\bar{\nabla}_{i}\partial_{j}\mathbf{A}_{\rho}-\partial_{\rho}\bar{\nabla}_{j}A_{i}\Big]\,,\\&=e^{-2A(\rho)}\partial_{\rho}\Big[\bar{\nabla}^{i}\partial_{i}\mathbf{\theta}-\bar{\nabla}_{i}\mathbf{A}^{i}\Big]=0\,.\label{eq:rhomaxwell}
    \end{split}
\end{equation}
Exerting the operator $e^{-(d-4)A(\rho)}\partial_{\rho} e^{dA(\rho)}$ on Equ.~(\ref{eq:rhomaxwell}), we get
\begin{equation}
    e^{2A(\rho)}\Big[\partial^{2}_{\rho}+(d-2)A'(\rho)\partial_{\rho}\Big]\Big[\bar{\nabla}^{i}\partial_{i}\mathbf{\theta}-\bar{\nabla}_{i}\mathbf{A}^{i}\Big]=0\,.
\end{equation}
Now we can decompose the bulk fields $\mathbf{A}_{i}(x,\rho)$ and $\mathbf{\theta}(x,\rho)$ into the eigenmodes of the differential operator $-e^{2A(\rho)}\Big[\partial^{2}_{\rho}+(d-2)A'(\rho)\partial_{\rho}\Big]$, and we obtain
\begin{equation}
m_{n}^{2}\Big[\bar{\nabla}^{i}\partial_{i}\mathbf{\theta}^{(n)}-\bar{\nabla}_{i}\mathbf{A}^{(n)i}\Big]=0\,,\label{eq:thetaeom}
\end{equation}
which are exactly the correct equations of motion for the tower of St\"{u}ckelberg/Goldstone boson fields $\theta^{(n)}(x)$. We note that this decomposition is consistent with the boundary condition Equ.~(\ref{eq:maxwellbc}) i.e. 
\begin{equation}
    \partial_{\rho}\theta(x,\rho)|_{\rho_{B}}=\mathbf{A}_{\rho}(x,\rho_{B})=0\,.
\end{equation}
Hence, in the dual island setup we have a tower of decoupled massive gauge fields in the AdS$_{d}$ as we expect. These gauge fields are normalizable KK modes of the bulk gauge field. From the bulk perspective, these KK modes are massive as the zero mass mode $\psi_{0}(\rho)=\text{const.}$ is not normalizable under the norm Equ.~(\ref{eq:normalization}). From the dual island setup perspective, these gauge fields correspond to a tower of nonlinearly realized U$(1)$ gauge symmetries.  

To verify the above understanding, we study the corresponding gauge symmetries. We have the bulk gauge transform
\begin{equation}
    \mathbf{A}_{\mu}(x,\rho)\rightarrow\mathbf{A}_{\mu}(x,\rho)+\partial_{\mu}\epsilon(x,\rho)\,.\label{eq:gaugetransform}
\end{equation}
We can decompose the bulk small gauge transformation parameter $\epsilon(x,\rho)$ into normalizable KK modes along the same line as the bulk gauge field as
\begin{equation}
    \epsilon(x,\rho)=\sum_{n=1}^{\infty}\psi_{n}(\rho)\epsilon^{(n)}(x)\,.\label{eq:epsilondecom}
\end{equation}
Furthermore, the large gauge transform corresponds to the zero mode of the bulk gauge parameter $\epsilon(x,\rho)$. In the dual  model on the branes, this bulk large gauge symmetry is a global symmetry and it is supported both by the bath and the AdS brane.

Thus, we have the gauge transforms for the gauge fields and the Goldstone bosons in the AdS$_{d}$ part of the dual island setup
\begin{equation}
    \bar{\mathbf{A}}^{(n)}_{i}(x,\rho)\rightarrow\bar{\mathbf{A}}^{(n)}_{i}(x,\rho)+\partial_{i}\epsilon^{(n)}(x,\rho)\,,\quad\theta^{(n)}(x)\rightarrow\theta^{(n)}(x)+\epsilon^{(n)}(x)\,,
\end{equation}
As a consistency check, the decomposition Equ.~(\ref{eq:epsilondecom}) for the gauge transformation parameter $\epsilon(x,\rho)$ preserves the boundary conditions Equ.~(\ref{eq:maxwellbc}) as it obeys
\begin{equation}
\partial_{\rho}\epsilon(x,\rho)|_{\rho=\rho_{B}}=0\,.
\end{equation}

Having understood the gauge symmetries in the dual island setup, let's study the global symmetry. The bath global symmetry charge operator can be deduced from the bulk using  Gauss' law as\footnote{We note that this is not the full global charge operator as it lacks a term that is supported on the defect. This operator duals to the charge of the bath global symmetry generated by $J_{2}^{a}$ in Sec.~\ref{sec:GlobalSymIsland}.}
\begin{equation}
    \hat{Q}_{\text{bath}}=\int d^{d-1}\vec{x}\sqrt{-\bar{g}}e^{(d-2)A(\rho_{c})} F_{0\rho}(t,\vec{x},\rho_{c})\,,\quad \rho_{c}\rightarrow\infty\,,
\end{equation}
which is also known to be consistent with the AdS/CFT dictionary.
In the AdS$_{d+1}$ bulk, by doing canonical quantization, we have the equal-time commutation relation
\begin{equation}
    [A_{\rho}(t,\vec{x},\rho),F_{0\rho}(t,\vec{y},\rho')]=i\frac{1}{\sqrt{-\bar{g}}e^{(d-2)A(\rho)}}\delta(\rho-\rho')\delta^{d-1}(\vec{x}-\vec{y})\,.
\end{equation} 
Thus we can see that the Goldstone bosons transform under this Global symmetry as\footnote{We note that $\int_{\rho_{B}}^{\infty} d\rho  e^{(d-4)A(\rho)}\psi_{n}(\rho)$ is finite for normalizable KK modes. This is because one can use Equ.~(\ref{eq:KKwave}) to show that that we have $\psi_{n}(\rho)\sim e^{-(d-2)A(\rho)}=\cosh^{-(d-2)}\rho$  as $\rho\rightarrow\infty$.}
\begin{equation}
[\hat{Q}_{\text{bath}},\theta(x,\rho)]=i\,\Rightarrow[\hat{Q}_{\text{bath}},\theta^{(n)}(x)]=i\int_{\rho_{B}}^{\infty} d\rho e^{(d-4)A(\rho)}\psi_{n}(\rho)\neq0\,,\label{eq:goldstonebosoncharge}
\end{equation} 
where we used $\theta^{(n)}(x)=\int_{\rho_{B}}^{\infty}d\rho e^{(d-4)A(\rho)}\psi_{n}(\rho)\theta(x,\rho)$.

In summary, in the dual island setup, we have a global symmetry that is supported on the bath and a tower of independent spontaneously broken U(1) gauge symmetries in the AdS. In the bulk, these symmetries assemble to a single U(1) gauge symmetry which is not spontaneously broken. Note that the matter fields in the dual island setup  in the Karch-Randall setup are strongly coupled which is distinct from the simple toy model with only a single gauge boson we considered in Sec.~\ref{sec:CFTanalysis}.  A scenario that is closer to the island setup we considered in Sec.~\ref{sec:GlobalSymIsland} would have a gauge boson in the gravitational AdS part whose mass is entirely due to  one-loop quantum corrections of the leaky matter fields. This corresponds more closely to the limit that the Karch-Randall brane is close to the asymptotic boundary, i.e. $\rho_{B}\rightarrow-\infty$. In this case, there is a distinct anomalously light KK mode, whose primary support is near the KR brane. This KK mode is the lightest normalizable mode $n=1$ of the bulk photon and its mass is scale separated from that of the other KK modes. We will further elaborate this point in a companion paper \cite{Geng:2026tba1}. To study the global symmetry in this model, we neglect the other KK modes so the truncated dual island setup is closer to the island setup we considered in Sec.~\ref{sec:GlobalSymIsland}.

\subsection{Operators, Global Symmetry Charges and the Extra Dimension}

In this subsection, we will see how the operators inside the island are charged under the preserved global symmetry and see that their global symmetry charges are encoded in the bath due to the fact that the Goldstone boson is dual to the bulk Wilson line along the $\rho$ direction.

Let's consider an operator $\hat{O}(x)$ which lives on the  KR brane and is charged under the gauge field $\bar{\mathbf{A}}_{i}^{(1)}(x)$, which is the lowest KK mode of the bulk photon that becomes anomalously light when $\rho_{B}\rightarrow-\infty$, as
\begin{equation}
    \hat{O}(x)\rightarrow e^{iq_{1}\epsilon^{(1)}(x)}\hat{O}(x)\,,
\end{equation}
then we can construct a gauge invariant operator using the associated Goldstone boson $\theta^{(1)}(x)$
\begin{equation}
    \hat{O}^{\text{Phys}}(x)=e^{-iq_{1}\theta^{(1)}(x)}\hat{O}(x)\,.\label{eq:branedresstheta}
\end{equation}
As we have discussed in Sec.~\ref{sec:GlobalSymIsland}, this dressing is necessary for operators inside the island. We will focus only on operators $\hat{O}(x)$ inside the island. This operator is charged under the global symmetry due to the global symmetry transform of the Goldstone boson $\theta^{(1)}(x)$ Equ.~(\ref{eq:goldstonebosoncharge}) as
\begin{equation}
    q_{\text{global}}=q_{1}\int_{\rho_{B}}^{\infty} d\rho e^{(d-4)A(\rho)}\psi_{1}(\rho)\,.\label{eq:qglobal}
\end{equation}

Interestingly, from Sec.~\ref{sec:globalgaugeKR}, the Goldstone boson $\theta^{(1)}(x)$ is a KK mode of the bulk Wilson line $\theta(x,\rho)$ that connects a bulk point $(x,\rho)$ to the boundary, i.e. the bath, through the extra dimension $\rho$. Thus, we can think of $\theta^{(1)}(x)$ as the associated KK component of the bulk Wilson line which connects the point $x$ on the AdS$_{d}$ part of the dual island setup to the point on the bath with the same coordinate $x$ (see Fig.\ref{pic:branethetadress}). This is thus the geometrization of the mechanism we uncovered in Sec.~\ref{sec:GlobalSymIsland} that the global symmetry charge of operators inside the island is encoded in the bath due to the Goldstone boson dressing.

 As a result, we can see that the global charge Equ.~(\ref{eq:qglobal}) of the operator in the island can be detected in the bath. This global symmetry charge is encoded in the bath due to the mixing between the global symmetry and the spontaneously broken gauge symmetries by the Goldstone boson and the existence of an extra dimension. This result is fully consistent with the results in analysis of the weakly coupled island setup in Sec.~\ref{sec:GlobalSymIsland} and the extra dimension provides a nice geometric picture for the mechanism that the global symmetry charges inside island are encoded in the bath.

\subsection{Gravitating Bath}
Before we wrap up the discussion in this section, let's make a quick remark on another consistency check between our construction of global symmetry and entanglement islands. 

\begin{figure}[h] 
\begin{centering}
\begin{tikzpicture}[scale=1]
\draw[-,very thick,black!100] (-2.5,0) to (0,0);
\draw[-,very thick,black!100] (0,0) to (2.5,0);
\draw[pattern=north west lines,pattern color=purple!200,draw=none] (0,0) to (-2.5,-1.875) to (-2.5,0) to (0,0);
\draw[pattern=north west lines,pattern color=purple!200,draw=none] (0,0) to (2.5,-1) to (2.5,0) to (0,0);
\draw[-,dashed,color=black!50] (0,0) to (-1.875,-2.5); 
\draw[-,dashed,color=black!50] (0,0) to (0,-2.875);
\draw[-,dashed,color=black!50] (0,0) to (1.875,-2.375); 
\draw[-,very thick,color=green!!50] (0,0) to (2.5,-1); 
\draw[-,very thick,color=green!!50] (0,0) to (-2.5,-1.875);
\node at (0,0) {\textcolor{red}{$\bullet$}};
\node at (0,0) {\textcolor{black}{$\circ$}};
\draw[-,dashed,color=black!50] (-2,-1.5) arc (-140:-21.8014095:2.5);
\draw[-,dashed,color=black!50] (-1.5,-1.125) arc (-140:-21.8014095:1.875);
\draw[-,dashed,color=black!50] (-1,-0.75) arc (-140:-21.8014095:1.25);
\end{tikzpicture}
\caption{\small A deformed setup of Karch-Randall braneworld with two branes. In the intermediate description, one has a gravitating bath.}
\label{pic:twobrane}
\end{centering}
\end{figure}
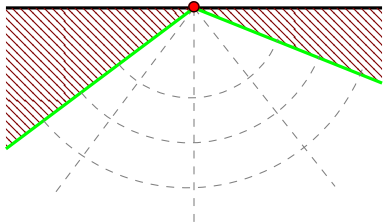

As it is discovered in \cite{Geng:2020fxl}, in a deformed setup of the Karch-Randall braneworld, there is no entanglement island. In this setup, one has two Karch-Randall branes as shown in Fig.~\ref{pic:twobrane}. In this case, the dual intermediate description is a modification of the island setup, where the bath is also gravitational.\footnote{This is the key reason for the absence of entanglement islands \cite{Geng:2020fxl}.} Thus, in this case the tension between global symmetry and the Bekenstein-Hawking entropy bound persists and we don't expect our construction of global symmetry charged operators inside the AdS would work in this case. To see that this is indeed the case, we notice that the zero mode of the KK wave equation Equ.~(\ref{eq:KKwave}) is now normalizable as the $\rho$ goes from the left to right brane in a finite range. Hence, in the intermediate description there is a massless gauge boson associated with this zero mode and we no longer have purely massive gauge bosons. In fact, in this case the intermediate theory is similar to the standard AdS/CFT scenario and the U(1) gauge symmetry associated with the massless gauge boson is dual to a global symmetry in a dual CFT. This dual CFT lives on the defect in Fig.~\ref{pic:twobrane}. This dual CFT ensures the Bekenstein-Hawking entropy bound which forbids the existence of global symmetry in the intermediate description.

\begin{figure}[h]
\begin{centering}
\begin{tikzpicture}[scale=1.4]
\draw[-,very thick,black!100] (-2.5,0) to (0,0);
\draw[-,very thick,black!100] (0,0) to (2.5,0);
\draw[pattern=north west lines,pattern color=purple!200,draw=none] (0,0) to (-2.5,-1.875) to (-2.5,0) to (0,0);
\draw[-,dashed,color=black!50] (0,0) to (-1.875,-2.5); 
\draw[-,dashed,color=black!50] (0,0) to (0,-2.875);
\draw[-,dashed,color=black!50] (0,0) to (1.875,-2.375); 
\draw[-,dashed,color=black!50] (0,0) to (2.5,-1.6875); 
\draw[-,very thick,color=green!!50] (0,0) to (-2.5,-1.875);
\node at (0,0) {\textcolor{red}{$\bullet$}};
\node at (0,0) {\textcolor{black}{$\circ$}};
\draw[-,dashed,color=black!50] (-2,-1.5) arc (-140:-2:2.5);
\draw[-,dashed,color=black!50] (-1.5,-1.125) arc (-140:-2:1.875);
\draw[-,dashed,color=black!50] (-1,-0.75) arc (-140:-2:1.25);
\node at (-1.25,-0.9375) {\textcolor{black}{$\cross$}};
\node at (-1.15,-1.3) {\textcolor{black}{$\hat{O}(x)$}};
\draw[-,very thick,snake it,color=black!50] (-1.25,-0.9375) arc (-135:-5:1.56);
\end{tikzpicture}
\caption{\small A demonstration of the dressing using the St\"uckelberg fields $\theta^{(n)}(x)$ for charged operators on the brane. The St\"{u}ckelberg fields $\theta^{(n)}(x)$ are the KK modes of the bulk Wilson line $\theta(x,\rho)$ i.e. $\theta^{(n)}(x)=\int_{\rho_{B}}^{\infty}e^{(d-4)A(\rho)}\psi_{n}(\rho)\theta(x,\rho)$. Therefore, $\theta^{(n)}(x)$ can be thought as a bulk Wilson line integrated against the KK wavefunction $\psi_{n}(\rho)$ with an end point at $(x,\rho_{B})$ on the brane and another end on the asymptotic boundary of the bulk. This smeared Wilson line goes purely in the $\rho$-direction.}
\label{pic:branethetadress}
\end{centering}
\end{figure}
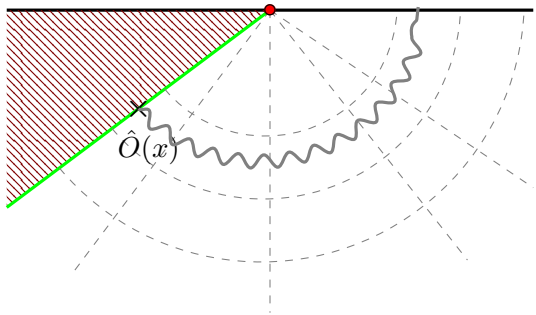

\section{Explicit Examples from String Theory}\label{sec:string}
In this section, we provide string theory constructions of the type of global symmetries we discussed in Sec.~\ref{sec:GlobalSymIsland} and Sec.~\ref{sec:KR}. We consider systems of D-branes that realize the Karch-Randall braneworld in the low-energy regime. We will see that the global symmetries we considered naturally exist in these setups. We will identify the various fields we had in Sec.~\ref{sec:KR} as the low-energy excitations in these setups.

We want to first remind the readers that in fully-fledged AdS/CFT, the isometry of the internal manifold, like $S^5$ in $AdS_{5}\times S^5$ for type IIB string on the background with D3-branes, is dual to the $R$ symmetry in the CFT, which is a global symmetry, that corresponds to a gauge symmetry in the AdS bulk with gauge fields given by the KK modes of the compact space, obtained from the full 10d or 11d graviton compactified on them. This point will be important for us to construct a setup that mimics the bottom-up setup we considered in the Karch-Randall braneworld in Sec.~\ref{sec:KR} in which the internal manifold $\mathbf{M}^6$ is six dimensional, i.e. the bulk geometry is of the form $AdS_{4}\times \mathbf{M}^{6}$ which deforms to $AdS_{5}\times S^5$ near the asymptotic boundary of the bulk. 

The Karch-Randall setup can be engineered in string theory \cite{Karch:2000ct,Karch:2001cw,DHoker:2007hhe,Aharony:2011yc,Bachas:2018zmb,Karch:2022rvr,Huertas:2023syg}. It consists of $D3$-branes in 10d Type IIB superstring theory ending on $D5$- and NS5-branes, preserving a quarter of the supersymmetries.\footnote{That is we have eight supercharges preserved.} The gravitational dual of this configuration is described by an asymptotic $AdS_5\times \mathbf{S}^5$ region (the bulk) connected to an $AdS_4\times \mathbf{M}^6$ geometry (which models the Karch-Randall brane) with $\mathbf{M}^6$ a manifold that smoothly caps off to the $\mathbf{S}^5$, see Fig.\ref{fig:bag_diagram}. In the following subsections we discuss its gauge and gravitational nature.

\subsection{The $D3$/$D5/NS5$ System in Type IIB String Theory}\label{sec:D3/D5/NS5}

The setup that we are interested in is 4d $\mathcal{N}=4$ $SU(N)$ SYM in a semi-infinite spacetime with a 3d boundary defect, which hosts a 3d $\mathcal{N}=4$ SCFT. This boundary is considered in \cite{Gaiotto:2008ak}, which can be described using the Hanany-Witten brane configurations \cite{Hanany:1996ie}. The 3d system can be coupled to the 4d system in different ways with the 3d super(conformal)symmetries preserved, realizing different half-BPS boundary conditions of the 4d $\mathcal{N}=4$ SYM. Note that much of this section is review of the string construction. We talk about the relevant global symmetries in Sec. \ref{TheGlobalSymmetries}.

This field theory setup is realized in string theory as follows. We start with a stack of $N$ $D3$-branes extending along the directions 0123. The gauge theory living on this stack of $D3$-branes is the usual 4d $\NN=4$ SYM $SU(N)$ theory. A boundary for this theory can be introduced by allowing the $D3$-branes to end on an appropriate configuration of NS5- and $D5$-branes, which are localized along the 3-direction and extend along 012 as well as additional transverse directions. To preserve the maximum supersymmetry, the NS5-branes are oriented along 012$\,$456 and coincide with the $D3$-branes in the directions 789, while the $D5$-branes extend along 012$\,$789 and share the same position as the $D3$-branes in the directions 456. All 5-branes are placed at the same location in the direction 3, thereby defining a boundary for the 4d $SU(N)$ theory. This boundary configuration is a Hanany-Witten \cite{Hanany:1996ie} setup of $D3$-branes suspended in the 3-direction among NS5- and $D5$-branes, describing a 3d $\NN=4$ CFT in the particular limit of all the 5-branes coinciding in the 3-direction.

As studied in \cite{Gaiotto:2008ak}, how the $D3$-branes end on the NS5- and $D5$-branes is very constrained. The upshot is that the NS5-branes (also the $D5$s) get grouped into $n$ stacks, with each stack containing $n_a$ branes associated with an irreducible $SU(2)$ representation of dimension $K_a$ \cite{Gaiotto:2008ak}. This means that each of the $n_a$ NS5-branes in the $a^{th}$ stack has $K_a$ $D3$-branes ending on it. The same applies to the $D5$-branes for which instead of $(n_{a},K_{a})$ we use $(m_{b},L_{b})$, leading to the constraint:
\begin{equation}\label{the_n_0}
N = \sum_a n_a K_a + \sum_b m_b L_b.
\end{equation}
where $N$ is the number of semi-infinite $D3$-branes. We will see that this constraint matches with the dual gravitational result in (\ref{the-n}), and this splitting has important consequences in the gravitational realization.

\subsection{The gravitational description}\label{sec:gravitydescription}

In this section, we describe the supergravity dual of the setup. \st{presented before.} It was constructed in \cite{DHoker:2007hhe,DHoker:2007zhm} (see also \cite{Aharony:2011yc}). We will see that the solution has two distinct regions: an $AdS_5 \times \IS^5$ region that is holographically dual to $\mathcal{N}=4$ SYM and an $AdS_4 \times \mathbf{M}^6$ region (where $M_6$ is a six-dimensional manifold we specify below) that is the dual of the 3d $\mathcal{N}=4$ SCFT with the interpolating region should describe how the dual 3d and 4d systems couple in the holographic dual. This $AdS_4$ region caps off the $AdS_5$ and thus describes the Karch-Randall (KR) brane. 

In order to describe the gravity solution, we should note that it has to be invariant under the 3d $\mathcal{N}=4$ superconformal group $OSp(4|4)$, whose bosonic symmetry is $SO(2,3) \times SO(3) \times SO(3)$. The overall structure of the metric is then:
\begin{equation}
ds^2 = f_4^2 ds^2_{AdS_4} + f_1^2 ds^2_{\IS^2_{(1)}} + f_2^2 ds^2_{\IS^2_{(2)}} + ds^2_\Sigma,\label{eq:the_metric}
\end{equation}
where $\Sigma$ is an oriented Riemann surface, which for our purpose will be taken to be the the second quadrant of the complex plane. The functions $f_1$, $f_2$, and $f_4$ depend on a complex coordinate $w=re^{i\varphi}$ (in our case $r\in (0,\infty)$ and $\varphi\in[\frac{\pi}{2},\pi]$) on $\Sigma$:
\begin{equation}
ds^2_\Sigma = 4 \rho^2 |dw|^2,
\end{equation}
where $\rho$ is a real function on $\Sigma$.

The metric can be obtained explicitly, using the following intermediate expressions \cite{DHoker:2007hhe}:

\beqa
& W\equiv \partial_w h_1\partial_{\bar w}h_2+\partial_w h_2\partial_{\bar w}h_1 
\; \;,\;\; N_1\equiv 2h_1h_2|\partial_w h_1|^2-h_1^2 W \;\; ,\;\;N_2\equiv 2h_1h_2|\partial_w h_2|^2-h_2^2 W\, .
\nonumber\\
\label{wnn}
\eeqa
The dilaton is given by
\beqa
e^{2\Phi}=\frac{N_2}{N_1}\, ,
\label{dilaton}
\eeqa
and the metric functions are given by
\beqa
\rho^2=e^{-\frac 12 \Phi}\frac{\sqrt{N_2|W|}}{h_1h_2}\; ,\; f_1^2=2e^{\frac 12\Phi} h_1^2\sqrt{\frac{|W|}{N_1}}\; ,\; f_2^2=2e^{-\frac 12\Phi} h_2^2\sqrt{\frac{|W|}{N_2}}\; ,\; f_4^2=2e^{-\frac 12\Phi} \sqrt{\frac{N_2}{|W|}}\nonumber\\
\label{the-fs}
\eeqa

We see that everything is constructed using the two harmonic functions $h_1$ and $h_2$ \cite{DHoker:2007hhe}, which depend on the coordinate $w$ on the Riemann surface $\Sigma$. The functions $h_1$ and $h_2$ are expressed as:
\begin{equation}\label{harmonic1}
h_1 = 4 \text{Im}(w) + 2 \sum_{b=1}^m \tilde{d}_b \log \left( \frac{|w + il_b|^2}{|w - il_b|^2} \right),
\end{equation}
\begin{equation}\label{harmonic2}
h_2 = -4 \text{Re}(w) - 2 \sum_{a=1}^n d_a \log \left( \frac{|w + k_a|^2}{|w - k_a|^2} \right).
\end{equation}
The parameters $k_a$ and $l_b$ are the location on $\Sigma$ of the different stacks of NS5 ($w=-k_{a}$) and $D5$-branes ($w=il_{b}$), which differ in their linking number, or the number of $D3$-branes that end on each of them. This number is given by
\beqa
K_a\equiv 32 \pi k_a\in\IZ \quad,\quad L_b\equiv 32\pi l_b\in\IZ\, .
\label{quant1}
\eeqa

Furthermore, the parameters $d_a$, ${\tilde d}_b$ are related to the total number of NS5- and $D5$-branes in each stack in the following way
\beqa
n_a\equiv 32\pi^2 d_a\in\IZ \quad ,\quad m_b\equiv 32 \pi^2 {\tilde d}_b\in\IZ\, .
\label{quant2}
\eeqa

The functions Equ.~(\ref{harmonic1}), Equ.~(\ref{harmonic2}) parameterize the warping of the geometry along $w$ and can describe an $AdS_5$ with a KR ETW brane. 
In the limit of large $|w|$ the metric Equ.~(\ref{eq:the_metric}) describes $AdS_5 \times S^5$ with $N$ units of RR 5-form flux, as can be checked using the explicit form of harmonic functions Equ.~(\ref{harmonic1}), Equ.~(\ref{harmonic2}) for $r\to\infty$ (see for example \cite{Aharony:2011yc}).

The number of $D3$-branes that backreact in the asymptotic AdS$_5\times\IS^5$, can be read from the metric, or it can be easily computed using Equ.~(\ref{quant1}) and Equ.~(\ref{quant2}), multipliying the number of 5-branes times the number of $D3$-branes that end on them:
\beqa
N=\sum_a n_a K_a+\sum_b m_b L_b \, .
\label{the-n}
\eeqa
which agrees with Equ.~(\ref{the_n_0}).

In the region of small $w$, the metric Equ.~(\ref{eq:the_metric}) describes a smooth closing of the 5-sphere, as the $D5$- and NS5-branes ``eat'' all the 5-form flux, allowing the 5-sphere to shrink, and the gravitational dual is $AdS_4 \times \mathbf{M}^6$. A diagram of the solution is presented in Fig.\ref{fig:bag_diagram}. 
\begin{figure}[h]
\centering
\includegraphics[scale=.15]{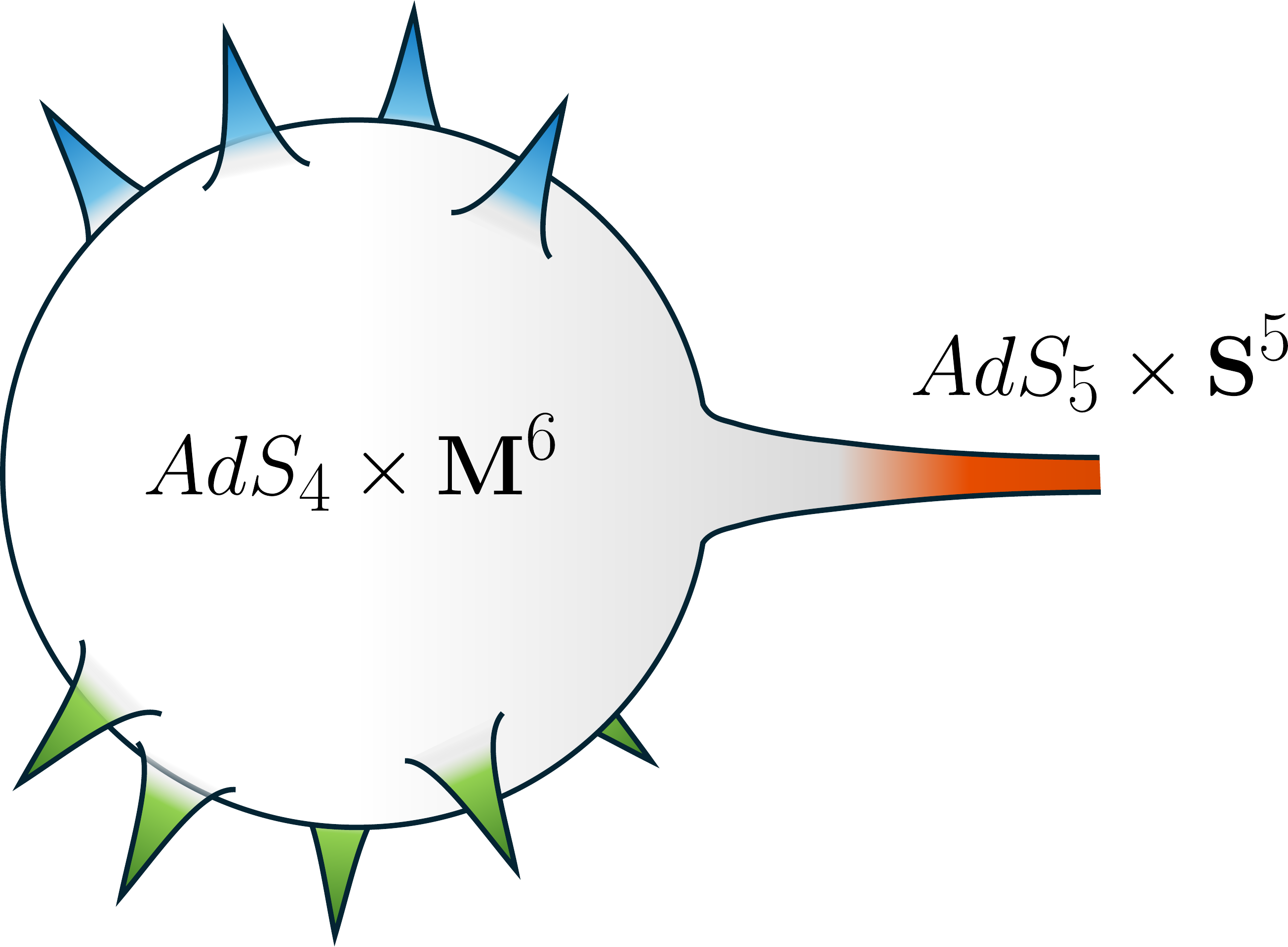}
\caption{\small Schematic diagram of the $AdS_4 \times \mathbf{M}^6$ ``fat'' Karch-Randall brane, with the blue and green spikes at the positions of the NS5- and $D5$-branes in the compact space. This bag has a throat attached, that asymptotes to $AdS_5\times S^5$, the $AdS_5$ ``bulk''.}
\label{fig:bag_diagram}
\end{figure}

We can see that the shape of this geometry is roughly the one of $AdS_4$ times a compact bag with some spikes at the positions of the 5-branes, which are given by $k_a$ and $l_a$. This bag has an $AdS_5\times S^5$ throat attached, which corresponds to the gravitational dual of the $D3$-branes ending on this web of 5-branes, see figure \ref{fig:bag_diagram}. This $AdS_4 \times \mathbf{M}^6$ is the ``fat'' Karch-Randall brane in which the $AdS_5\times S^5$ ends. Notice that this string theory construction doesn't have scale separation, so the length scales of the compact spaces are the same as their corresponding $AdS$ regions (see for example \cite{Bachas:2011xa,Bachas:2018zmb,Anastasi:2025puv}. Moreover, this supergravity solution is not by itself UV complete due to the spikes \cite{Assel:2011xz}. These spikes have a nice resolution in string theory as they should be replaced by the corresponding $5$-branes and the singularities just depict the fact that there are extra localized light degrees of freedom at these spikes that are captured by the low-energy worldvolume dynamics of the $5$-branes. Since the $D5$-branes and the NS5-branes are related to each other by $S$-duality, we choose to focus on the $D3$/$D5$ cases.

\subsection{The Global Symmetries}\label{TheGlobalSymmetries}

We now want to provide realizations of the type of global symmetries we discussed in Sec.~\ref{sec:GlobalSymIsland} and Sec.~\ref{sec:KR} in this string theory setup. The basic considerations and constructions are aligned with the bottom-up study in Sec.~\ref{sec:KR}. Namely, a gauge symmetry in the supergravity background Equ.~(\ref{eq:the_metric}) should be dual to a global symmetry in the dual 4d $\mathcal{N}=4$ SYM with 3d $\mathcal{N}=4$ SCFT boundary and in the effective intermediate description such a global symmetry will be mixed with spontaneously broken gauge symmetry and there exist  operators in the  AdS$_{4}$ that are charged under this global symmetry.  Thus, the goal of this section is to construct/identify gauge symmetries consistent with the background Equ.~(\ref{eq:the_metric}).

In the model we considered in Sec.~\ref{sec:GlobalSymIsland}, we coupled a global symmetry in the bath to a global symmetry on its boundary through a marginal deformation Equ.~(\ref{eq:CFTdescription}), which is continuously tunable. However, in this string theory setup, as the solution is fully backreacted, the gravitational coupling between the ETW (the ``bag'' in figure \label{fig:bag_diagram}) and the bulk (the $AdS_5$ throat in figure \label{fig:bag_diagram}) is fixed by the number and positions of branes (see for instance \cite{Bachas:2018zmb,Demulder:2022aij,Huertas:2023syg}), it is not a marginal coupling, and we expect that the gauge symmetries in the bag and the throat will be coupled with the same strength as the gravity part. Therefore, weak coupling or weak leakage from the ETW part, will be given only by the scarcity of the bulk $CFT_4$ degrees of freedom, with respect those of the boundary $CFT_3$. This is a limit that we can obtain by increasing the number of 5-branes with respect to the semi-infinite D3-branes.

Nevertheless, apart from this remark about the strength of the coupling, and the fact that we have 5 extra compact dimensions without scale separation, the rest of the analysis is similar to the previous sections.

Having said that, we now want to figure out what are the candidates for the gauge symmetry in this setup. We would want them to arise naturally from the string theory solution.

The simplest way to obtain a gauge symmetry in the background Equ.~(\ref{eq:the_metric}) is to introduce additional spacetime filling branes in this background, for example a stack of M D7-branes,\footnote{This stack of D7-branes fills the AdS$_{4}$ and the Riemann surface $\Sigma$  with the other two directions wrapped on an $S^{2}$.} as considered in \cite{Karch:2002sh} (also used for the same purpose in \cite{Anastasi:2025puv}). If M is much smaller than N, we are in the probe limit, so we can take the approximation that the D7-branes don't backreact on the background Equ.~(\ref{eq:the_metric}). Therefore, in the field theory description, we introduced a fundamental flavor group $U(M)$, with its global currents which are dual to the gauge fields living on the worldvolume of the spacetime filling D7-branes. This is a straightforward way to introduce a gauge field in the bulk of our string theory solution and is the realization in this setup of the toy model we considered in Sec.~\ref{sec:KR}.  

The above example clearly has the properties of the symmetries in Sections 3 and 4. Here we study a different symmetry with a nice field theory interpretation. This is the $R$ symmetry of the 4d $\mathcal{N}=4$ SYM with 3d $\mathcal{N}=4$ SCFT boundary. In the field theory description, this global symmetry transforms different supercharges to each other. In the system we are considering, the original $SO(6)$ $R$ symmetry of the 4d $\mathcal{N}=4$ SYM is broken down to $SO(3)\times SO(3)$, which is the $R$ symmetry of 3d $\mathcal{N}=4$ superconformal algebra, due to the coupling to the $3d$ system on its boundary. This consideration suggests that the global symmetry analogous to that used in Sec.~\ref{sec:GlobalSymIsland} should be the $SO(3)\times SO(3)$ $R$ symmetry that has support on both the 4d system and its 3d boundary. The current of the $SO(3)\times SO(3)\subset SO(6)$ of the 4d system is not conserved by itself due to the coupling to the $3d$ system. Neither is the current of the $SO(3)\times SO(3)$ of the 3d system by itself. Namely, the 4d system and the 3d system are exchanging the $SO(3)\times SO(3)$ charges such that their total $SO(3)\times SO(3)$ charge is conserved. In the dual gravity description, this R-symmetry corresponds to the isometries of the compact dimensions, i.e. those of the two $S^{2}$'s in Equ.~(\ref{eq:the_metric}), which also form $SO(3)\times SO(3)$. Interestingly, in the asymptotic regime far from the bag,\footnote{Remember that the bag gives an effective Karch-Randall brane.} the two $S^{2}$'s combine to form an $S^{5}$. The $S^{5}$ has the isometry $SO(6)$, and the asymptotic geometry is $AdS_5\times \mathbf{S}^5$. This behavior of the compact directions can be interpreted as that the D3/D5 system, which is captured by the bag, breaks the compact isometries of the original $AdS_5\times \mathbf{S}^5$ from $SO(6)$ down to $SO(3)\times SO(3)$,\footnote{The non-compact isometries are dual to the conformal symmetries on the field theory side.} and this is in analogy to the R-symmetry breaking pattern in the field theory description. In fact, in the bulk the $SO(3)\times SO(3)$ is a genuine gauge symmetry with the associated gauge fields. The gauge fields are components of the 10d metric with one index along the two $S^{2}$'s as in the usual Kaluza-Klein compactification.

In summary, we provided two types of realizations of the island setup global symmetry in the string theory setup we depicted in Sec.~\ref{sec:D3/D5/NS5} and Sec.~\ref{sec:gravitydescription}. This setup, as a potential string theory uplift of the Karch-Randall braneworld, has been used to provide explicit constructions of entanglement islands in string theory \cite{Uhlemann:2021nhu,Demulder:2022aij,Karch:2022rvr}. These realizations of the global symmetry are aligned with our bottom-up studies in Sec.~\ref{sec:GlobalSymIsland} and Sec.~\ref{sec:KR}. Many interesting calculations regarding these realizations can be done on both the field theory side and the supergravity/string theory side, and we will come to this study in the near future.

\section{Resolving the Harlow-Shaghoulian Puzzle}\label{sec:resolvingHS}
With various concrete constructions of global symmetries in the island setup and a clear understanding of why global symmetry is consistent with entanglement islands and how it is encoded in the bath, we are ready to discuss the puzzle proposed by Harlow and Shaghoulian in \cite{Harlow:2020bee}. We first review the Harlow-Shaghoulian puzzle and then provide the resolution of it based on our understanding of global symmetries in the island setup in Sec.~\ref{sec:GlobalSymIsland} and Sec.~\ref{sec:KR}.

\subsection{The Harlow-Shaghoulian Puzzle}
Harlow and Shaghoulian argued in \cite{Harlow:2020bee} that a global symmetry is not consistent with entanglement islands. More specifically, they argued that there cannot exist operators inside the entanglement island that are charged under a global symmetry. Their argument is based on the following consideration.\footnote{See \cite{Harlow:2018tng,Harlow:2018jwu,Bao:2024hwy} for relevant considerations in different contexts and \cite{Heckman:2024oot} for some comments on these considerations.} Let's take an operator $\hat{O}_{q}^{\mathcal{I}}(x)$ with charge $q$ under a global symmetry inside the entanglement island $\mathcal{I}$ in a $d$-dimensional island setup. By the holographic duality, the global symmetry is supported in the bath and the CFT$_{d-1}$ and by the holographic interpretation of entanglement islands, the global symmetry of the operator $\hat{O}_{q}^{\mathcal{I}}(x)$ should be encoded in the bath subregion $R$. Then due to the locality in the non-gravitational bath, one can write the symmetry transformation operator in the union of the bath and the CFT$_{d-1}$ as
\begin{equation}
    U(g)=U_{\bar{R}}(g)\Pi_{i}U_{R_{i}}U_{\text{edge}_{i}}(g)\,,
\end{equation}
where $\bar{R}$ denotes the complement of the bath subregion $R$ and it contains the CFT$_{d-1}$, $R$ is decomposed as the union of the small subregions $R_{i}$ with the codimension two boundary between them denoted as $\text{edge}_{i}$.\footnote{Note that we take $R$ and $\bar{R}$ to be a subregion on a Cauchy slice so $R$ and $\bar{R}$ themselves are codimension one.}
Since $\hat{O}_{q}^{\mathcal{I}}(x)$ is claimed to be fully encoded in $R$, its global symmetry transform is purely implemented by
\begin{equation}
U_{R}(g)=\Pi_{i}U_{R_{i}}(g)U_{\text{edge}_{i}}(g)\,.\label{eq:split}
\end{equation}
However, an apparent contradiction comes if one takes each $R_{i}$ to be small enough so that they themselves don't have an island, i.e. $EW(R_{i})=R_{i}$. This is because, by the entanglement wedge reconstruction, operators inside the entanglement wedge should commute with operators outside of it. Thus, one might conclude that each $U(R_{i})$ commutes with $\hat{O}_{q}^{\mathcal{I}}(x)$ and so does $U_{\text{edge}_{i}}$ as the edges are one dimension lower. This would imply that their multiplication, i.e. $U_{R}$, will also commute with $\hat{O}_{q}^{\mathcal{I}}(x)$ and this contradicts the assumption that $\hat{O}_{q}^{\mathcal{I}}(x)$ is charged under the global symmetry.

\subsection{The Resolution of the Harlow-Shaghoulian Puzzle}
The resolution of the above puzzle is clear given our understanding that the global symmetry is always mixed with a bulk gauge symmetry and the fact that the AdS Goldstone boson is non-locally supported in both the AdS and in the bath. In the island setup, the operators inside the island are dressed using the Goldstone boson, so these dressed composite operators are gauge-invariant and have support also in the bath. Hence, the operator $\hat{O}_{q}^{\mathcal{I}}(x)$ localized inside the island from the perspective of the AdS necessarily intersects the entanglement wedge of a small bath subregion $R_{i}$. Thus, $U_{R_{i}}(g)$ can detect it. Note that the island operator $\hat{O}^{\mathcal{I}}_{q}(x)$ carries a lot of information and its bath dual is highly nonlocal and is not necessarily splittable as Equ.~(\ref{eq:split}). We are suggesting that only the information of its global charge is splittable in the bath. Thus, the operator $\hat{O}^{\mathcal{I}}_{q}(x)$ doesn't have to be inside the entanglement wedge of the small subregion $R_{i}$. Hence, there is no contradiction between such a global symmetry and the entanglement wedge reconstruction. This is readily seen in the Karch-Randall braneworld where it is clear that from the bulk perspective the global symmetry is in fact gauged and the Goldstone boson dressing is equivalent to a bulk smeared Wilson line dressing from the brane to the asymptotic boundary through the extra dimension. For the non-holographic island setups, one could think of the dressing to the bath as through a quantum extra dimension and the Goldstone boson is a microscopic wormhole operator \cite{Geng:2025rov}.

\section{Conclusions and Discussions}\label{sec:conclusion}
In this paper, we provided concrete evidence for the conjecture that the absence of global symmetry in quantum gravity is tied to unitarity and the realization of holography. Our strategy was to provide explicit constructions of global symmetries in controlled setups in which unitarity is violated. The examples we gave correspond to island setups that contain a gravitational asymptotically AdS spacetime coupled to a nongravitational bath on its asymptotic boundary. In such scenarios, in which the AdS contains an entanglement island,  the  AdS gravitational theory is not unitary. The holographic interpretation of entanglement islands states that information inside the islands is encoded in subregions in the nongravitational bath. Without an island, the who gravitational AdS region would be captured by its boundary, which however does not contain enough states to allow for a global symmetry in the AdS. On the other hand,  black holes in island setups are no longer captured by the quantum system on its boundary but by the bath subregion that claims the island. The entanglement island overlaps the black hole interior. This resolves the apparent tension between a global symmetry and the Bekenstein-Hawking entropy formula. To put this conceptual understanding on firm ground, we provided explicit constructions of global symmetries in the island setup from both the bottom-up perspective and the top-down perspective. In these constructions, the previous understanding that gauge fields in the AdS are massive due to the bath induced spontaneous symmetry breaking in the island setup \cite{Geng:2020qvw,Geng:2020fxl,Geng:2021hlu,Geng:2023zhq,Geng:2023ynk,Geng:2025rov} played an essential role. General lessons from these constructions are that the global symmetry is always mixed with a spontaneously broken gauge symmetry and the charged operators in AdS under the global symmetry that are detectable by the bath can  exist  only inside the island and they are in fact nonlocally supported in both the AdS and the bath from the point of view of the full island setup. As a bonus, we realized a resolution of the Harlow-Shaghoulian puzzle \cite{Harlow:2020bee} from the above general lessons. Last but not least, we should notice that of entanglement islands, which is essential for our construction of global symmetry, universally exist in the island setup.

\section*{Acknowledgments}
We thank Safi Bahcall, Ning Bao,  Liam A. Fitzpatrick, Jonathan Heckman, Luca Iliesiu, Yikun Jiang, Juan Maldacena, Yasunori Nomura, Matt Reece, Edgar Shaghoulian, Huy Tran and Ángel Uranga for helpful discussions. 
HG, LR and DT are supported by the Gravity, Spacetime, and Particle Physics (GRASP) Initiative from Harvard University. JH is supported by a Simons Foundation's fellowship through the Targeted Grant to Instituto Balseiro. AK is supported in part  by DOE grant DE-SC0022021 and by a grant from the Simons Foundation (Grant 651678, AK).

\bibliographystyle{JHEP}

\bibliography{main}
\end{document}